\documentclass[11pt]{article}
\usepackage{comment}
\usepackage[margin=2.5cm]{geometry}
\usepackage{amsmath,amssymb,extarrows,mathtools,graphicx,subfigure,setspace,bbold}
\usepackage{cite}
\usepackage{slashed}
\usepackage[dvipsnames]{xcolor}
\usepackage{hyperref}
\hypersetup{colorlinks=true, linkcolor=blue, citecolor=magenta, linktoc=page}
\usepackage{tikz}
\usepackage{arydshln,leftidx,mathtools}
\usetikzlibrary{decorations.pathreplacing}
\numberwithin{equation}{section}

\newcommand{\mcal}{\mathcal}

\DeclareMathOperator{\Tr}{Tr}

\def\({\left(}
\def\){\right)}

\newcommand{\be}{\begin{equation}}
\newcommand{\ba}{\begin{eqnarray}}
\newcommand{\ea}{\end{eqnarray}}
\newcommand{\ee}{\end{equation}}

\newcommand{\f}{\frac}

 \def\f {\frac}

\newcommand{\bal}{\begin{align}}
\newcommand{\eal}{\end{align}}
\newcommand{\bes}{\begin{equation*}}
\newcommand{\bas}{\begin{eqnarray*}}
\newcommand{\eas}{\end{eqnarray*}}
\newcommand{\ees}{\end{equation*}}
\newcommand{\bals}{\begin{align*}}
\newcommand{\eals}{\end{align*}}
\newcommand{\nn}{\nonumber}
\newcommand{\p}{\partial}

\begin{document}

\begin{titlepage}
\thispagestyle{empty}

\begin{flushright}
IPM/P-2015/040\\
\end{flushright}

\vspace{.4cm}
\begin{center}
\noindent{\Large \textbf{On the Entanglement Between Interacting Scalar Field Theories}}\\
\vspace{2cm}

M. Reza Mohammadi Mozaffar
and
Ali Mollabashi
\vspace{1cm}

{\it School of Physics,\\Institute for Research in Fundamental Sciences (IPM),\\Tehran, Iran}\\
\vspace{1cm}
{\it Email:} m$_{-}$mohammadi@ipm.ir , mollabashi@ipm.ir

\vskip 2em
\end{center}

\vspace{.5cm}
\begin{abstract}
We study ``field space entanglement'' in certain quantum field theories consisting of $N$ number of free scalar fields interacting  with each other via kinetic mixing terms. We present exact analytic expressions for entanglement and Renyi entropies between arbitrary numbers of scalar fields by which we could explore certain entanglement inequalities. Other entanglement measures such as mutual information and entanglement negativity have also been studied. We also give some comments about possible holographic realizations of such models.
\end{abstract}

\end{titlepage}

\newpage

\tableofcontents
\noindent
\hrulefill


\onehalfspacing
\section{Introduction}
Quantum entanglement offers different measures to capture some non-local properties in quantum field theories (QFTs). There are various measures for quantum entanglement including entanglement and Renyi entropies \cite{Renyi} which measure the amount of quantum entanglement between various parts of the Hilbert space of the theory. Among these measures, specifically entanglement entropy (EE) has recently gained a huge amount of interest.

In this context, the most common way available in the literature for studying quantum entanglement is based on a one-to-one correspondence between localized degrees of freedom of local quantum field theories and plane waves as a particular complete basis spanning their total Hilbert space. Based on such a map the Hilbert space is decomposed as $\mcal{H}=\mcal{H}_A\otimes\mcal{H}_B$, where $A$ and $B$ correspond to spatial subregions such that $\mcal{M}=A\cup B$ is a constant time slice of the manifold which the QFT is defined on. Such a decomposition is reliable up to the spatial resolution introduced by the UV cut-off of the theory. The spatial subregions $A$ and $B$ are defined via a co-dimension-two surface $\p A$. Following such a decomposition and tracing out either part $A$ or $B$ leads to a measure for the quantum entanglement between localized degrees of freedom in spatial regions $A$ and $B$. We denote this type of EE as ``spatial entanglement entropy" (SEE). Some well known features of entanglement entropy such as the celebrated area-law divergence \cite{Bombelli:1986rw, Srednicki:1993im} is peculiar to SEE.

SEE is not the only type of EE one can define between various degrees of freedom of a single field. There are other types of EE corresponding to different Hilbert space decompositions. For example one can decompose a given Hilbert space into states with specific energies and consider the EE referring to given scale of energy $\Lambda$. This type of EE is known as ``momentum space entanglement entropy" which measures the EE between degrees of freedom of a single field below and above a given energy scale $\Lambda$ in the momentum space (see e.g. \cite{Balasubramanian:2011wt}).\footnote{There are also two other types of entanglement discussed in the literature: The first one which is called ``entanglement in theory space" is defined via gauging (un-gauging) two theories with global symmetries in \cite{Yamazaki:2013xva}. We would like to thank Mukund Rangamani for bringing our attention to this reference. The other one which is called ``global symmetry entanglement" is defined via partitioning the symmetry group in \cite{Taylor:2015kda}.}

\vspace*{4mm}

If more than one field lives in a field theory, one may ask about probable entanglement between degrees of freedom corresponding to different fields. In contrast to various EE measures defined between different degrees of freedom of a single field, the entanglement between degrees of freedom of different fields is caused via possible interactions between them.\footnote{We are aware of some studies which can be considered as quantum mechanical counterparts of such an analysis, including reference \cite{Rangamani:2015qwa} where entanglement between non-interacting qubits is studied and also reference \cite{Haque} where a particle partitioning is considered for studying entanglement entropy.}
Using the terminology of reference \cite{Taylor:2015kda}, we denote this type of EE as ``field space entanglement entropy" (FSEE).   

It is worth to note that Ryu-Takayanagi proposal \cite{RT, Ryu:2006ef, Nishioka:2009un, Takayanagi:2012kg} for holographic entanglement entropy is by construction a proposal to compute SEE in a field theory which supports classical Einstein theory as a gravity dual. A natural question which may arise is about the possibility of a holographic realization for other types of EE e.g. FSEE. We are not going to answer this question in this paper and we will only give some comments about it in the section \ref{sec:dis}. Recently some arguments about this interesting question has appeared in the literature specifically in \cite{Mollabashi:2014qfa} and \cite{Taylor:2015kda} (see also \cite{Karch:2014pma} for some related holographic improvements). 

\vspace*{4mm}

In this paper we try to further investigate the notion of FSEE from a field theoretic point of view. To do so we consider various field theories which are interacting with each other. The interaction between these field theories is responsible for generating entanglement between them. In order to study the entanglement between these theories we integrate out a generic number of them which leads to a reduced density matrix. Next we follow the standard procedure to study entanglement and Renyi entropies.

For simplicity we focus on scalar field theories with Gaussian interactions between them. Since such models are Gaussian, they are analytically tractable to a satisfactory extent, and thus we consider them as a simple laboratory to study some general properties of FSEE. Explicitly we work out the generic reduced density matrix of such models and study entanglement and also all Renyi entropies analytically. A similar construction have been previously studied in \cite{Mollabashi:2014qfa} and in the context of condensed matter physics in \cite{Xu:2011gn, Furukawa:2010nd, Chen:2013kba, Lundgren:2013}. 

The authors of reference \cite{Mollabashi:2014qfa} have considered two free scalar field theories denoted by $\phi$ and $\psi$ which interact homogeneously in a $d$-dimensional space-time via two types of interactions: kinetic mixing (marginal) and massive interactions. They have decomposed the total Hilbert space of the theory as $\mcal{H}=\mcal{H}_\phi\otimes\mcal{H}_\psi$ and integrated out the states in $\mcal{H}_\psi$ and worked out the entanglement and Renyi entropies between $\phi$ and $\psi$ in the ground state which is no more a direct product due to the interaction between them.

In this paper we generalize the procedure of reference \cite{Mollabashi:2014qfa} in the sense that we consider $N$ free field theories defined on a common $d$-dimensional flat space-time which interact with each other. The action is thus given by
\be
S=\int dx^d \left[\mcal{L}_1\left(\phi_1\right)+\mcal{L}_2\left(\phi_2\right)+\cdots+\mcal{L}_N\left(\phi_N\right)+\mcal{L}_{\mathrm{int.}}\left(\phi_i\right)\right],
\ee
where $\mcal{L}_i\left(\phi_i\right)$ with $i=1,2,\cdots,N$ denote the Lagrangian density of free field theories and $\mcal{L}_{\mathrm{int.}}\left(\phi_i\right)$ denotes all possible interactions between them. We are interested in entanglement and Renyi entropies between these field theories which is generated via the interaction term $\mcal{L}_{\mathrm{int.}}\left(\phi_i\right)$. The total Hilbert space of this model can be decomposed as
$$\mathcal{H}_{\mathrm{tot.}}=\mathcal{H}_1\otimes\mathcal{H}_2\otimes\cdots\otimes\mathcal{H}_N,$$
where $\mathcal{H}_i$'s denote the Hilbert space of each field theory defined by $\mcal{L}_i\left(\phi_i\right)$. We are interested in the entanglement between generic $m$ number of these field theories with the rest $(N-m)$ of them. To do so we consider the following more compact notation for the decomposition of the total Hilbert space as
\be
\mathcal{H}_{\mathrm{tot.}}=\mathcal{H}_{(m)}\otimes\mathcal{H}_{(N-m)}
\ee  
where $\mathcal{H}_{(m)}$ is defined as $\mathcal{H}_{(m)}=\mathcal{H}_1\otimes\mathcal{H}_2\otimes\cdots\otimes\mathcal{H}_m$ and $\mathcal{H}_{(N-m)}$ similarly denotes the Hilbert space of the rest $(N-m)$ field theories. In such a way we define the reduced density matrix $\rho_{(m)}$ by tracing out the $\mathcal{H}_{(N-m)}$ part of the Hilbert space
$$\rho_{(m)}=\mathrm{Tr}_{\mathcal{H}_{(N-m)}}\left[\rho_{\mathrm{tot.}}\right],$$
which leads to the following definition of entanglement and Renyi entropies
\begin{align}\label{eq:EERE}
S_{\mathrm{ent.}}(m)=-\mathrm{Tr}\left[\rho_{(m)}\log\rho_{(m)}\right]\;\;\;\;\;,\;\;\;\;\;
S^{(n)}(m)=\frac{1}{1-n}\log\mathrm{Tr}\left[\rho_{(m)}^n\right].
\end{align}

The rest of this paper is organized as follows: In section 2 we introduce two different models called ``infinite-range" and ``nearest-neighbour" models which are different in the range of their interactions. In section 3 we report the results of calculating the reduced density matrix of generic number of fields and compute entanglement and Renyi entropies of these two models. In section 4 we investigate different features of these models probing them by entanglement measures including entanglement inequalities and $n$-partite information.
In the discussion section we will give some comments about the holographic dual of such a construction and also a field theoretic counterpart for black-hole evaporation process. In appendix \ref{sec:app1} we explain some details related to the calculation of the reduced density matrix of our models. 

\section{Kinetic Mixing Gaussian Models}\label{sec:KMGM}
In this paper we are interested in Gaussian models as the simplest examples of interacting field theories which are analytically tractable. The most general wave functional for such models is given by \cite{Callan:1994py}
\begin{align}\label{eq:WF}
\Psi[\phi_i] =\mcal{N} \exp \Bigg\{-\frac{1}{2}\int dx^{d-1} & dy^{d-1}\sum_{i,j=1}^N\phi_i(x)G_{ij}(x,y)\phi_j(y)\Bigg\},
\end{align}
where $\mcal{N}$ is a normalization constant and $G_{ij}(x,y)$'s are complex valued functions which are symmetric on $i,j$ indices and also on the variables $x$ and $y$. The corresponding (total) density matrix is constructed as $\rho_{\mathrm{tot.}}[\phi'_i,\phi_i]=\Psi^*[\phi'_i]\Psi[\phi_i]$. One can define a generic reduced density matrix by integrating out (without loss of generality) the first $m$ number of the fields on the whole space-time as
\be\label{eq:rhom}
\rho_{(m)}[\phi'_{m+1},\phi_{m+1},\cdots,\phi'_{N},\phi_{N}]=\int\mathcal{D}\phi_1\cdots\mathcal{D}\phi_{N-m}\Psi^*[\phi'_i]\Psi[\phi_i],
\ee
where $\left(\phi'_1,\cdots,\phi'_m\right)$ is identified with $\left(\phi_1,\cdots,\phi_m\right)$ in the integrand.

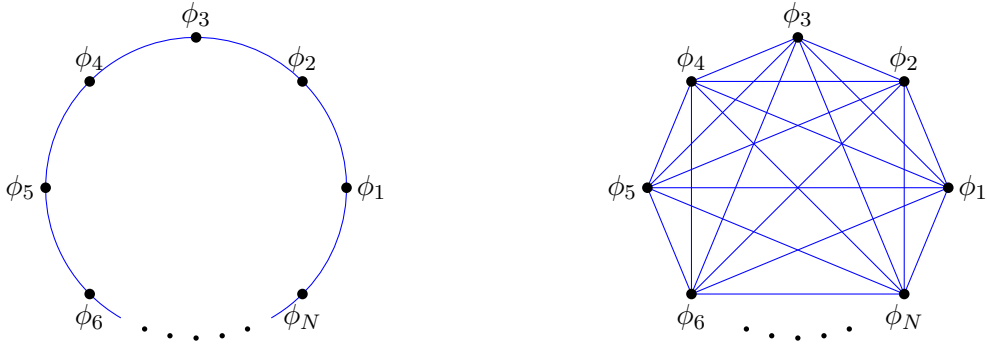
\begin{figure}
\begin{center}
\begin{tikzpicture}[>=latex]
\draw [color=blue](1, -1.732) arc (-60:240: 2cm and 2cm);
\draw (0, 0)            coordinate      (fov)
           +(0:2cm)  coordinate (A)  (fov)
           +(45:2cm)  coordinate (B)  (fov)
           +(90:2cm)  coordinate (C)  (fov)
           +(135:2cm)  coordinate (D)  (fov)
           +(180:2cm)  coordinate (E)  (fov)
           +(225:2cm)  coordinate (F)  (fov)
           +(315:2cm) coordinate (H);
\draw (0, 0)            coordinate      (fov)
           +(250:2cm)  coordinate (G1)  (fov)
           +(260:2cm)  coordinate (G2)  (fov)
           +(270:2cm)  coordinate (G3)  (fov)
           +(280:2cm)  coordinate (G4)  (fov)
                    +(290:2cm)  coordinate (G5);
    \fill[radius=2pt]
   (A)   circle node[anchor=west] {$\phi_1$}
   (B)   circle node[anchor=south] {$\phi_2$}
      (C)   circle node[anchor=south] {$\phi_3$}
   (D)   circle node[anchor=south] {$\phi_4$}
   (E)   circle node[anchor=east] {$\phi_5$}
      (F)   circle node[anchor=north] {$\phi_6$}
   (H)   circle node[anchor=north] {$\phi_N$};
    \fill[radius=1pt]
      (G1)   circle node {}
      (G2)   circle node {}
      (G3)   circle node {}
      (G4)   circle node {}
      (G5)   circle node {};
\draw (8, 0)            coordinate      (fov)
           +(0:2cm)  coordinate (A)  (fov)
           +(45:2cm)  coordinate (B)  (fov)
           +(90:2cm)  coordinate (C)  (fov)
           +(135:2cm)  coordinate (D)  (fov)
           +(180:2cm)  coordinate (E)  (fov)
           +(225:2cm)  coordinate (F)  (fov)
           +(315:2cm) coordinate (H);
\draw (8, 0)            coordinate      (fov)
           +(250:2cm)  coordinate (G1)  (fov)
           +(260:2cm)  coordinate (G2)  (fov)
           +(270:2cm)  coordinate (G3)  (fov)
           +(280:2cm)  coordinate (G4)  (fov)
                    +(290:2cm)  coordinate (G5);
\draw [blue] (A) -- (B);
\draw [blue] (A) -- (C);
\draw [blue] (A) -- (D);
\draw [blue] (A) -- (E);
\draw [blue] (A) -- (F);
\draw [blue] (A) -- (H);
\draw [blue] (B) -- (C);
\draw [blue] (B) -- (D);
\draw [blue] (B) -- (E);
\draw [blue] (B) -- (F);
\draw [blue] (B) -- (H);
\draw [blue] (C) -- (D);
\draw [blue] (C) -- (E);
\draw [blue] (C) -- (F);
\draw [blue] (C) -- (H);
\draw [blue] (D) -- (E);
\draw [blue] (D) -- (F);
\draw [blue] (D) -- (H);
\draw [blue] (E) -- (F);
\draw [blue] (E) -- (H);
\draw [blue] (F) -- (H);
    \fill[radius=2pt]
   (A)   circle node[anchor=west] {$\phi_1$}
   (B)   circle node[anchor=south] {$\phi_2$}
      (C)   circle node[anchor=south] {$\phi_3$}
   (D)   circle node[anchor=south] {$\phi_4$}
   (E)   circle node[anchor=east] {$\phi_5$}
      (F)   circle node[anchor=north] {$\phi_6$}
   (H)   circle node[anchor=north] {$\phi_N$};
    \fill[radius=1pt]
      (G1)   circle node {}
      (G2)   circle node {}
      (G3)   circle node {}
      (G4)   circle node {}
      (G5)   circle node {};
\end{tikzpicture} 
\caption{Schematic plots of infinite-range model (right) and nearest-neighbour model (left) in the field space. The blue lines connecting different fields represents the existence of an interaction term between the corresponding fields. The infinite-range model clearly has much more interactions than the nearest-neighbour model.}
\label{Fig:Models}
\end{center}
\end{figure}

Since we are interested in analytically tractable simple models, in what follows we have chosen the same value of coupling constant between our mutually interacting field theories which means all off-diagonal non-vanishing elements of $G_{ij}$ take the same value. We are mainly interested in two models that we define in the following subsections. In the first model, any $\phi_i$ interacts with all other fields $\phi_j$ with $(i\neq j)$. This model is called infinite-range model.\footnote{This terminology is borrowed from the literature of statistical physics where e.g. an Ising model which all sites interact with each other are called infinite-range interacting Ising model. Here ``range'' refers to the field space rather than the real space. We thank Ali Naji for introducing us with this terminology. We also thank Julien Vidal for bringing our attention to related  references \cite{JVidal}, where aspects of a closely related model, knows as the Lipkin-Meshkov-Glick (LMG) model, has been  studied previously.}
In our second model any field $\phi_i$ interacts only with its nearest neighbours which are $\phi_{i\pm1}$. We consider this model with a periodic boundary condition in the field space and call it the nearest-neighbour model. See Fig.\ref{Fig:Models} for a geometric realization of these models in the field space.

Since we are interested in Gaussian models, in both of our models we consider kinetic mixing terms as the interaction between the free scalar fields, thus we are always dealing with marginal couplings. Note that both of these models in the special case where the total number of fields is two $(N=2)$ reduce to the massless interaction model in \cite{Mollabashi:2014qfa}.
\subsection{Infinite-Range Model}\label{sec:model1}
The infinite-range model is defined by the following action
\be\label{eq:action1}
S=\frac{1}{2}\int d^dx \left[\sum_{i=1}^N\left(\p_\mu\phi_i\right)^2+\lambda\sum_{i<j\leq N}^N\p_\mu\phi_i\p^\mu\phi_j\right],
\ee
where all $\phi_i$'s are interacting mutually with the same coupling constant $\lambda$. The wave functional of this model is given by Eq.\eqref{eq:WF} where
\be\label{eq:Gij1}
G(x,y)=\frac{W(x,y)}{2}
\begin{pmatrix}
            2 & \lambda & \lambda & \cdots &\lambda   \\
            \lambda &2&\lambda&\cdots &\lambda \\
             \lambda&\lambda&2&\cdots &\lambda \\
            \vdots &\vdots &\vdots &\ddots &\vdots \\
             \lambda&\lambda&\lambda&\cdots &2 \\
             \end{pmatrix},
\ee
and $W(x,y)=V^{-1}\sum_k|k|e^{ik(x-y)}$. We have briefly explained some details of this model in appendix \ref{sec:app1}.

One can easily show that this model can be diagonalized with the following eigenvalues
\be\label{eq:A1}
A_\alpha=1-\frac{\lambda}{2}\;\;\;\;,\;\;\;\;\alpha=1,2,\cdots,N-1\;\;\;\;\;\;,\;\;\;\;\;\;\;A_N=1+(N-1)\frac{\lambda}{2},
\ee
and after the corresponding orthogonal transformation one can rewrite this model in terms of new (primed) degrees of freedom
\be\label{eq:Sdiag}
S=\frac{1}{2}\int d^dx \sum_{i=1}^NA_i\left(\p_\mu\phi'_i\right)^2.
\ee
It is an easily task to check that the positivity of the Hamiltonian restricts the value of $\lambda$ to the following window
\be\label{eq:posI}
-\frac{2}{N-1}< \lambda < 2,
\ee
which we will consider in what follows as the range where this model is well-defined. This model is shown schematically in the field space in the right part of Fig.\ref{Fig:Models}.
\subsection{Nearest-Neighbour Model}
The nearest-neighbour model is defined with the following action
\be\label{eq:action2}
S=\frac{1}{2}\int d^dx \left[\sum_{i=1}^N\left(\p_\mu\phi_i\right)^2+\lambda\sum_{\langle i,j\rangle}\p_\mu\phi_i\p^\mu\phi_j\right],
\ee
where $\langle i,j\rangle$ means that the summation runs over two neighbours of each $\phi_i$ which are $\phi_{i\pm1}$. Because of symmetry considerations we impose a periodic boundary condition (in the field space) such that the nearest neighbours of $\phi_1$ are $\phi_2$ and $\phi_N$. It is obvious that the number of interactions in this model is much less than the infinite-range model. The wave functional of this model is also given by Eq.\eqref{eq:WF} where
\be\label{eq:Gij2}
G(x,y)=\frac{W(x,y)}{2}
\begin{pmatrix}
            2 & \lambda & 0 & \cdots &0 &\lambda   \\
            \lambda &2&\lambda&\cdots &0&0 \\
             0&\lambda &2&\cdots &0&0 \\
            \vdots &\vdots &\vdots &\ddots &  \vdots &\vdots \\
            0&0&0&\cdots &2&\lambda \\
             \lambda&0&0&\cdots &\lambda&2 \\
             \end{pmatrix},
\ee
and again $W(x,y)=V^{-1}\sum_k|k|e^{ik(x-y)}$ (see appendix \ref{sec:app1}).

One can easily show that the nearest-neighbour model can also be diagonalized and expressed in terms of new (primed) free fields just as Eq.\eqref{eq:Sdiag}. The eigenvalues of $G$ for the case of $N=2$ is
\be A_{1,2}=1\mp \frac{\lambda}{2},\ee
and for the case of $N(>2)$ is
\begin{align}
\begin{split}
\begin{cases}
A_{1,N}=1\mp \lambda,\;A_{2,3}=1\mp \lambda \cos\f{2\pi}{N},\dots, A_{N-2,N-1}=1\mp \lambda \cos\f{(N-2)\pi}{N} & ~~ N\mathrm{: even}  \\
A_1=1+\lambda,\;A_{2,3}=1- \lambda \cos\f{\pi}{N},\dots, A_{N-1,N}=1- \lambda \cos\f{(N-2)\pi}{N} & ~~ N \mathrm{: odd} 
\end{cases}
\end{split}
\end{align}
After performing the orthogonal transformation which leads to Eq.\eqref{eq:Sdiag}, one can compute the Hamiltonian of this model and show that the positivity of the Hamiltonian restricts the value of $\lambda$ to the following windows
\begin{align}\label{eq:posII}
\begin{split}
\begin{cases}
-2< \lambda < 2 & ~~ N\mathrm{: 2}  \\
-1< \lambda < 1 & ~~ N\mathrm{: even}  \\
-1< \lambda < \left(\cos\f{\pi}{N}\right)^{-1} & ~~ N\mathrm{: odd} 
\end{cases}
\end{split}
\end{align}
In what follows we consider the above range for the coupling constant $\lambda$ where this model is well-defined. The schematic plot of the nearest-neighbour model is given in the left part of Fig.\ref{Fig:Models}.
\section{Entanglement and Renyi Entropies}
In this section we report the results of computing the reduced density matrix and hence entanglement and Renyi entropies in our models using replica trick. Here we skip the details of the messy calculations leading to $\mathrm{Tr}[\rho_{(m)}^n]$, and we just present the final results. The interested reader may find some details about the essential steps of the computations in appendix \ref{sec:app1}.
\subsection{Infinite-Range Model}
Considering the infinite-range model one can show that using the definition of the reduced density matrix $\rho_{(m)}$ given in Eq.\eqref{eq:rhom}, together with the standard method of replication one can calculate $\mathrm{Tr}[\rho_{(m)}^n]$ which leads to (see appendix \ref{sec:app1} for details)
\be\label{eq:TrrhonI}
\mathrm{Tr}\left[\rho_{(m)}^n\right]=\mathcal{N}\prod_{i}\prod_{r=1}^n\left[1+f(m,N)\cos\left(\frac{2\pi r}{n}\right)\right]=\mathcal{N}\prod_{i}\frac{(1-\xi_i^n)^2}{(1+\xi_i^2)^n},
\ee
where
\be
f(m,N)=\frac{4(N-m)Y(m)}{4(N-m)Y(m)+(N-m)\lambda+2-\lambda}\;\;\;,\;\;\;Y(m)=-\frac{1}{4}\left(\frac{\lambda}{2}\right)^2\cdot\frac{2m}{2+(m-1)\lambda},
\ee
and $\xi_i$ is
\be
f(m,N)=\frac{2\xi_i}{1+\xi_i^2}.
\ee
Note that the normalization constant $\mathcal{N}$ plays no role in entanglement and Renyi entropies thus we will ignore it in what follows. Also note that in what follows we drop the index $i$ of $\xi_i$ which regards to the discretized real space since all $\xi_i$'s have the same value denoted by $\xi$.

Since the $m$ traced out fields together with the rest $(N-m)$ fields build up the whole system (the total density matrix corresponds to a pure state), one would expect the above expression to be invariant under $m\to(N-m)$ which is manifest in the expression of $f(m,N)$.  

\begin{figure}
\begin{center}
\includegraphics[scale=.68]{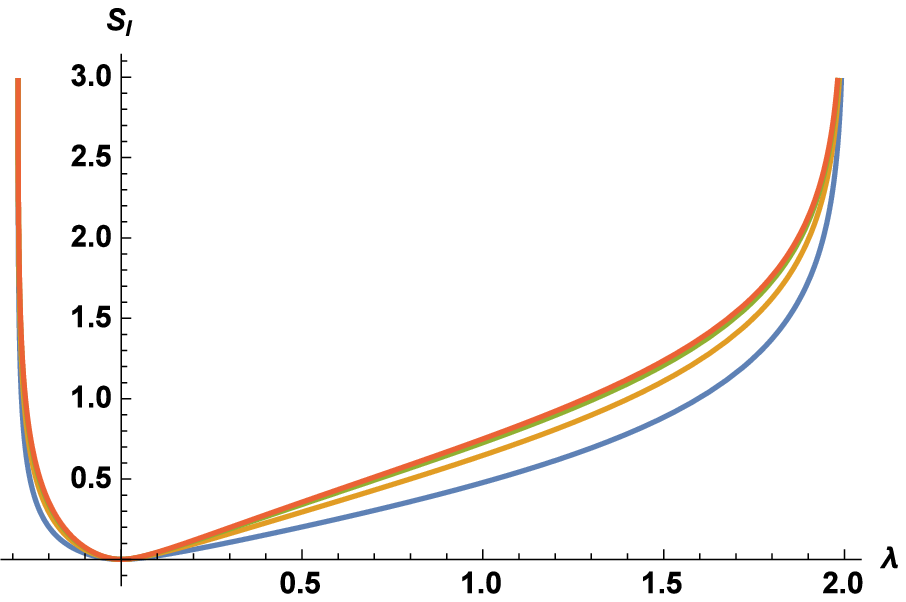}
\hspace{1cm}
\includegraphics[scale=.68]{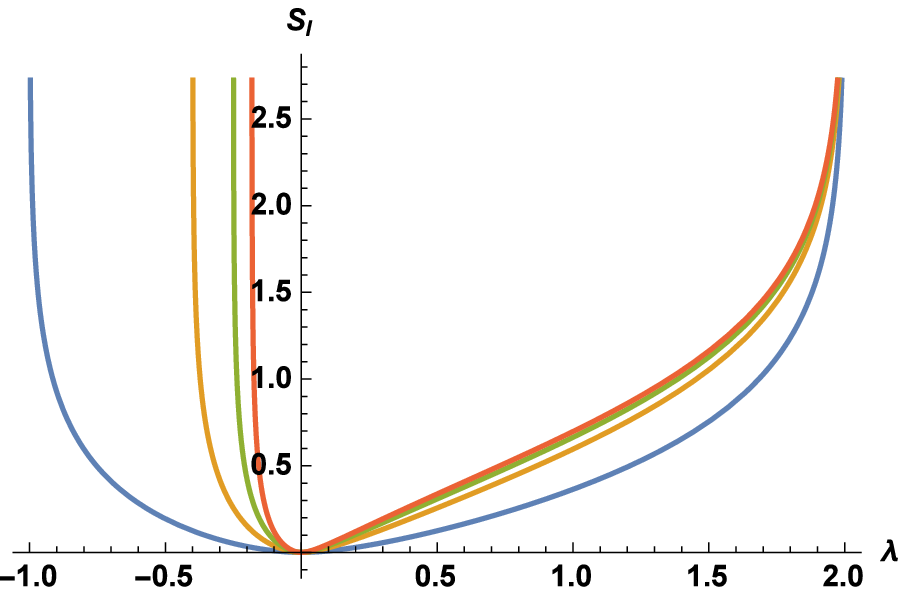}
\end{center}
\caption{Entanglement entropy of the infinite-range model as a function of coupling $\lambda$. 
}
\label{fig:EEMIlambda}
\end{figure}

Now we are equipped with everything needed to apply the definitions given in Eq.\eqref{eq:EERE} for entanglement and Renyi entropies. One can read the entropies as
\begin{align}
S^{(n)}&\equiv \sum_i s^{(n)}(\xi)=s^{(n)}(\xi) \sum_{\vec{k}\neq0} 1,\;\;\;\;S \equiv \sum_i s(\xi)=s(\xi) \sum_{\vec{k}\neq0}1,\nonumber\\
s^{(n)}(\xi)&=\frac{n\ln (1-\xi)-\ln(1-\xi^n)}{1-n},\;\;\;\; s(\xi)= \left[- \ln (1-\xi)-\dfrac{\xi}{1-\xi} \ln \xi \right],
\label{eq:general entropy}
\end{align}
where the infinite sum is UV divergent. In order to regularize these expressions we use a smooth momentum cut-off, i.e., $e^{-\epsilon|k|}$. If we consider the $(d-1)$-dimensional spatial manifold to be a $(d-1)$-torus with size $L$, the infinite sum simplifies to
\begin{align}\label{eq:uvcutoff}
\sum_{\vec{k}\neq0}1\sim\left(\sum_{k\neq0}e^{-\epsilon|k|}\right)^{d-1}=c_{d,d-1}\left(\frac{L}{\epsilon}\right)^{d-1}+c_{d,d-2}\left(\frac{L}{\epsilon}\right)^{d-2}+\cdots+c_{d,0},
\end{align}
where $c_i$'s are constants that only depends on $d$. All the terms of the resultant entanglement entropy are divergent and depend on the UV cut-off $\epsilon$ except the last one which is a universal term. To investigate the physical features of this model\footnote{This argument is also valid for the nearest-neighbour model.} in the following sections we will consider this universal term which is proportional to $c_{d,0}$. Also note that according to Eq.\eqref{eq:general entropy} the whole $\lambda$-dependence of entropies in this model is carried by $s^{(n)}(\lambda)$ and $s( \lambda)$. See Fig.\ref{fig:EEMIlambda} where the universal part of entanglement entropy of this model is plotted for different values of $m$ and $N$. 

Since in this paper we are dealing with entanglement in the field space, in what follows by entanglement and Renyi entropies we mean the ``density" of these quantities which is defined as the entanglement and Renyi entropies in units of the infinite volume factor Eq.\eqref{eq:uvcutoff}. This is obviously also true for the case of other entanglement measures which we define in the following including mutual and tripartite information. Thus here we have constructed the entanglement measures to be finite by definition. This is different from what happens in the case of spatial entanglement entropy. In that case some entanglement measures e.g. mutual information is defined by the whole expression of entanglement entropy which includes an area divergence but the divergent terms cancel out as long as the entangling regions do not have an intersection.\footnote{Although in the case of spatial entanglement entropy it is well known that tripartite information and in general $n$-partite information with $n>2$ are UV finite quantities even when the entangling regions share boundaries, this seems not to be generally correct as there is a counter example with corner shape entangling regions which have a single point as a common boundary \cite{CCEM}.} 

\subsection{Nearest-Neighbour Model}
Next we consider the nearest-neighbour model which again by using the definition of the reduced density matrix given in Eq.\eqref{eq:rhom} together with the standard method of replication we calculate $\mathrm{Tr}\left[\rho_{(m)}^n\right]$ for $m$ \textit{neighbour} fields out of $N$ ones which leads to the following results
\begin{align}\label{eq:TrrhonII}
\begin{split}
\mathrm{Tr}\left[\rho^n(m,N)\right]=\mathcal{N}\prod_i \prod_{r=1}^n&\left[1+\frac{2Y_-(m)g_-(N-m-1)}{2Y_-(m)g_-(N-m-1)-g_-(N-m+1)}\cos\left(\frac{2\pi r}{n}\right)\right]\times\\
&\prod_{s=1}^n\left[1+\frac{2Y_+(m)g_+(N-m-1)}{2Y_+(m)g_+(N-m-1)+g_+(N-m+1)}\cos\left(\frac{2\pi s}{n}\right)\right],
\end{split}
\end{align}
where
\begin{alignat}{2}\label{eq:TrrhonII2}
g_\pm(N)&=\prod_{s=1}^{\frac{N}{2}}\left[1-\lambda\,\cos\left(\frac{d_\pm(N)+2s}{N}\pi\right)\right],  &\hspace{1cm} d_\pm(N)&=\sin^2\left(\frac{2N+1\mp 1}{4}\pi\right)-1,\nn\\
Y(m)&=-\frac{1}{4}\left(-\frac{\lambda}{2}\right)^{m+1}\cdot\frac{1}{Z(m+1)}, &\hspace{1cm}
Y_d(m)&=-\frac{1}{4}\left(-\frac{\lambda}{2}\right)^2\cdot\frac{Z(m)}{Z(m+1)},\\
Y_\pm(m)&=Y(m)\pm Y_d(m), &\hspace{1cm}
Z(m)&=\prod_{r=1}^{m-1}\left[1-\lambda\cos\left(\frac{r}{m}\pi\right)\right]\nn.
\end{alignat}
\begin{figure}
\begin{center}
\includegraphics[scale=.68]{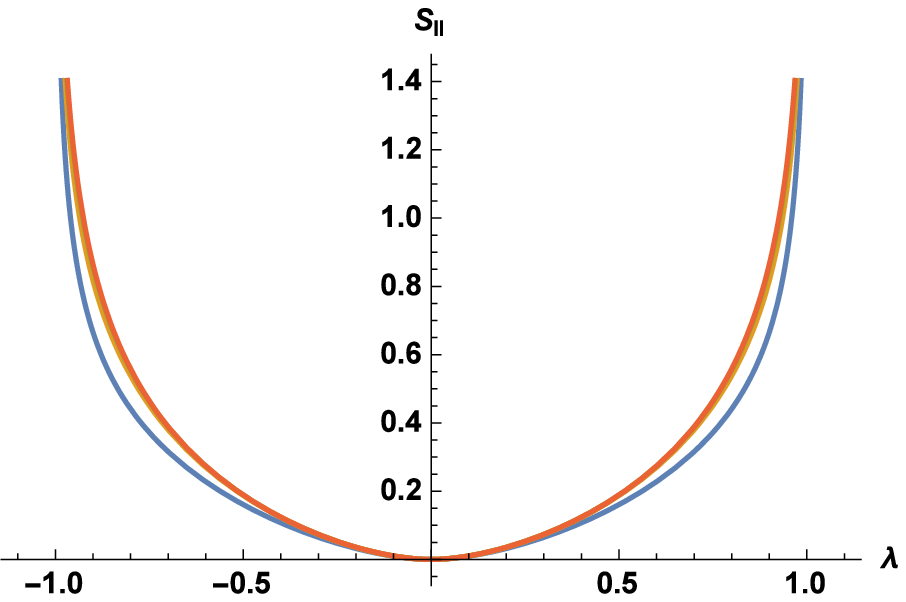}
\hspace{1cm}
\includegraphics[scale=.68]{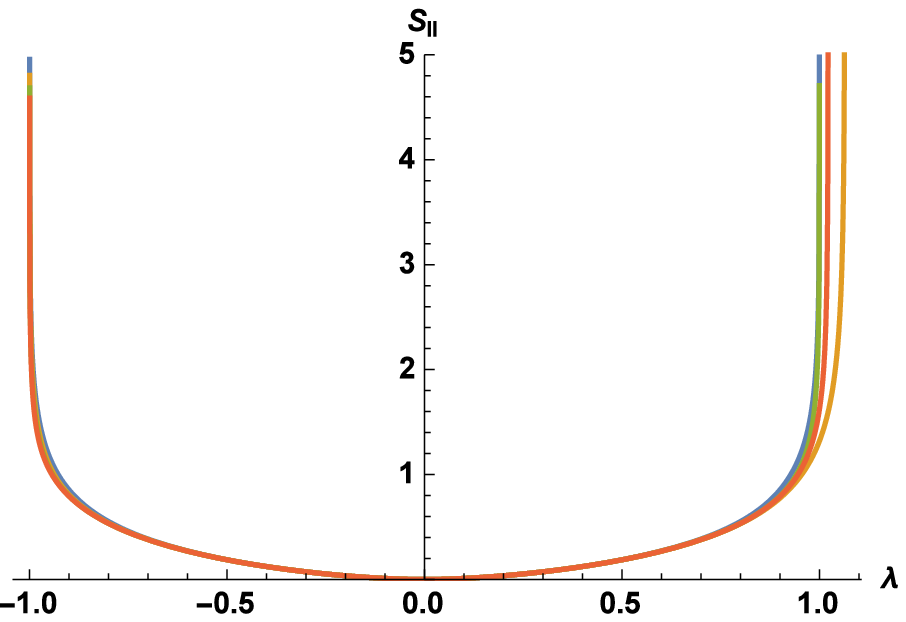}
\end{center}
\caption{Entanglement entropy for nearest-neighbour model as a function of coupling $\lambda$.}
\label{fig:EEMIIlambda}
\end{figure}

We define $g_+(0)\equiv\frac{1}{2}$ for consistency with the infinite-range model in the case of $N=2$.\footnote{Note that the infinite-range and nearest-neighbour models are the same for the case of $N=2$ and $N=3$.}
Again $\mathcal{N}$ is irrelevant to the calculation of entanglement and Renyi entropies. The result Eq.\eqref{eq:TrrhonII} is valid for $1\leq m<N-1$.
For the case of $m=N-1$ one should use the following expression
\be\label{eq:mmN}
\mathrm{Tr}\left[\rho^n(m=N-1,N)\right]=\mathcal{N}\prod_{r=1}^n\left[1+\frac{2\tilde{Y}_d(m)}{2\tilde{Y}_d(m)+1}\cos\left(\frac{2\pi r}{n}\right)\right]\;\;\;\;\;,\;\;\;\;\;\tilde{Y}_d(m)=-\frac{\lambda^2}{16}\frac{g_+(N-2)}{g_+(N)}
\ee
which of course is equal to the result of $m=1$ from Eq.\eqref{eq:TrrhonII} as expected. It is not hard to show that one can sum up the results of Eq.\eqref{eq:TrrhonII} and Eq.\eqref{eq:mmN} in a single formula as
\be\label{eq:NN2}
\mathrm{Tr}\left[\rho^n(m,N)\right]=\left(\mathrm{Tr}\left[\rho^n(m,N)\right]\mathrm{Tr}\left[\rho^n(N-m,N)\right]\right)^\frac{1}{2}
\ee
which is valid for $1\leq m<N$. The advantage of using this more compact formula is two-fold: it is no longer a piecewise formula and also the $m\to N-m$ symmetry becomes manifest in this form. Mathematically there is no difference between using Eq.\eqref{eq:TrrhonII} together with Eq.\eqref{eq:mmN}, or Eq.\eqref{eq:NN2}. In what follows we will continue with the first choice.
 
The expressions for the entanglement and Renyi entropies are similar to the infinite-range model given in Eq.\eqref{eq:general entropy}, and we just have to replace $s^{(n)}(\xi)$ and $s(\xi)$ with $s^{(n)}(\xi_+)+s^{(n)}(\xi_-)$ and $s(\xi_+)+s(\xi_-)$ respectively where $\xi_\pm$ are solutions of
\begin{align}
\frac{2\xi_{\pm}}{1+\xi_{\pm}^2}=\frac{2Y_{\pm}(m)g_{\pm}(N-m-1)}{2Y_{\pm}(m)g_{\pm}(N-m-1)\pm g_{\pm}(N-m+1)}.
\end{align}
For the case of $m=N-1$ we consider $s^{(n)}(\tilde{\xi})$ and $s(\tilde{\xi})$ where $\tilde{\xi}$ is defined as $\tilde{\xi}=\frac{2\tilde{Y}_d(m)}{2\tilde{Y}_d(m)+1}$. Finally note that as we have mentioned before, the structure of the regularization is independent of the interaction terms, thus in this model it exactly obeys the same structure of the previous model given in Eq.\eqref{eq:uvcutoff}.

\begin{figure}
\begin{center}
\includegraphics[scale=.7]{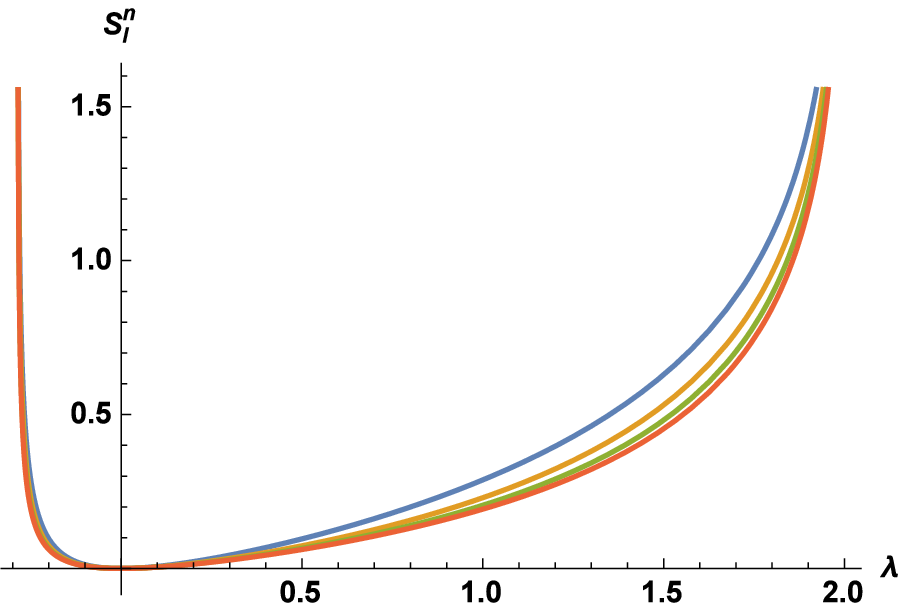}
\includegraphics[scale=.7]{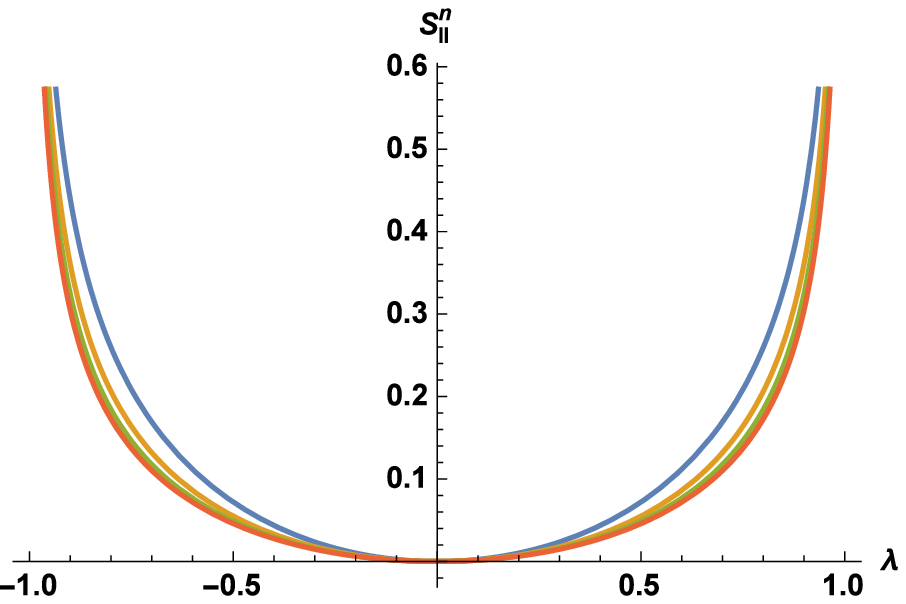}
\end{center}
\caption{Renyi entropies for the infinite-range model (left) and nearest neighbour model (right).}
\label{fig:RElambda}
\end{figure}

\section{Aspects of Field Space Entanglement}
In this section we investigate some important features of these models based on the entanglement measures computed in the previous section. First we discuss about some features of entanglement and Renyi entropies of these two models. Next we study some physical constraints on entanglement measures which are known as entanglement inequalities. We also study $n$-partite information for certain values of $n$, and entanglement negativity as two other entanglement probes in our models. This analysis may be helpful to gain a more physical intuition about the structure of entanglement in these models and perhaps more generally some generic physical features of field space entanglement. 

\subsection{Infinite-Range Versus Nearest-Neighbour Model}
In this subsection we are going to compare the infinite-range and the nearest-neighbour models using some graphical analysis. Previously in Fig.\ref{fig:EEMIlambda} and Fig.\ref{fig:EEMIIlambda} we have plotted the entanglement entropy of these two models as a function of the coupling constant $\lambda$. Note that the Hamiltonian positivity condition for these models which was given in Eq.\eqref{eq:posI} and Eq.\eqref{eq:posII}, results in a $N$-dependence for the valid range of coupling $\lambda$. This has caused some asymmetries in the entanglement and Renyi entropies under $\lambda\rightarrow -\lambda$. Also note that in the case of $\lambda=0$, since the vacuum state of the these models reduces to a direct product state, there is no entanglement between the specified degrees of freedom in these models. Fig.\ref{fig:RElambda} shows the Renyi entropy for these models as a function of coupling $\lambda$ for various Renyi indices $n$. These plots clearly show that $S^{(n)}_{\rm{I,II}}$ is a decreasing function of $n$ as expected. 

\begin{figure}
\begin{center}
\includegraphics[scale=.59]{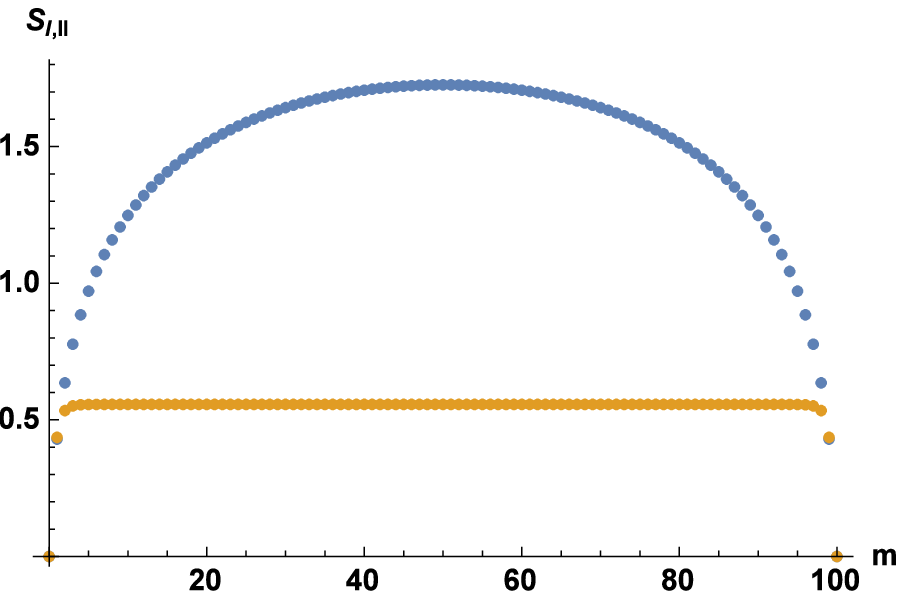}
\includegraphics[scale=.59]{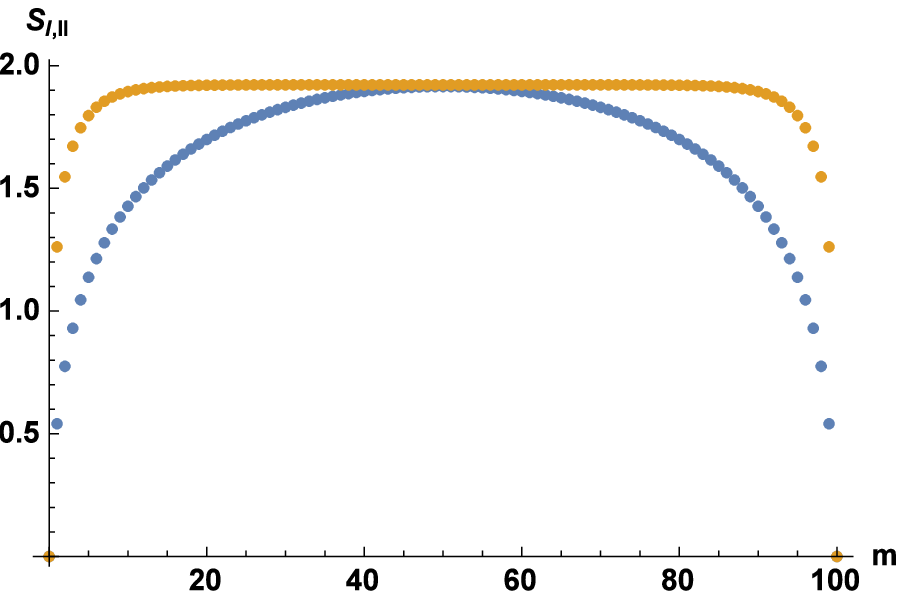}
\includegraphics[scale=.59]{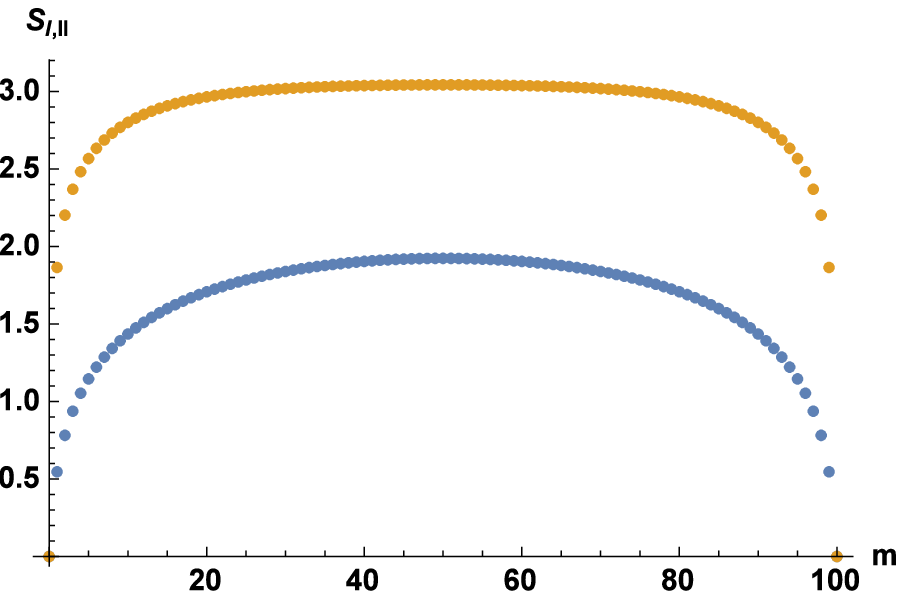}
\end{center}
\caption{Entanglement entropy of infinite-range (blue) and nearest-neighbour (orange) models for different values of coupling. From left to right: $\lambda=0.8$, $\lambda=0.99=\lambda_*$, $\lambda=0.999$ and we have set $N=100$.}
\label{fig:EEMI}
\end{figure}

In Fig.\ref{fig:EEMI} we have demonstrated the $m$-dependence of the EE in these two models for three different values of $\lambda$. Considering the coupling constant $\lambda$, the domain of validity of the infinite-range model is wider than the nearest-neighbour model (compare Eq.\eqref{eq:posI} and Eq.\eqref{eq:posII}). As the value of $\lambda$ starts increasing from $\lambda=0$, for $N>3$ which the distinction between these two models makes sense, the nearest-neighbour model reaches its maximum value of coupling constant, which we call $\lambda^{\mathrm{II}}_{\mathrm{max}}$, before the infinite-range one $(\lambda^{\mathrm{II}}_{\mathrm{max}}<\lambda^{\mathrm{I}}_{\mathrm{max}})$. Since as $\lambda\to\lambda^{\mathrm{I,II}}_{\mathrm{max}}$ the maximum value of the corresponding EE diverges, the value of the EE for the nearest-neighbour model starts to grow much faster than the infinite-range one as $\lambda\to\lambda^{\mathrm{II}}_{\mathrm{max}}$. Therefore there always exists a $\lambda_*(<\lambda^{\mathrm{II}}_{\mathrm{max}}<\lambda^{\mathrm{I}}_{\mathrm{max}})$ where the value of the EE of the nearest-neighbour model touches the value of that of the infinite-range one and gets larger values for $\lambda>\lambda_*$.

It is also interesting to study Renyi entropy as a function of Renyi index $n$. This is done in Fig.\ref{fig:renyi-n} where we have plotted the Renyi entropy (normalized by entanglement entropy) in our models for various parameter values as a function of $n$. In this figure the dashed black curve corresponds to the value of entanglement entropy which coincides at $n=1$ with Renyi entropy at arbitrary coupling $\lambda$. There exists two other interesting limits of Renyi entropy corresponding to $n\rightarrow 0$ and $n\rightarrow \infty$. In the $n\rightarrow 0$ limit, one can easily check that Renyi entropy by definition, Eq.\eqref{eq:EERE}, reduces to the Hartley entropy
\begin{align}
S^{(0)}=\lim_{n\rightarrow 0} S^{(n)}=\log \mathcal{D},
\end{align}
where $\mathcal{D}$ is the dimension of the image of the reduced density matrix. Since in our models $\mathcal{D}$ is infinite, as it can be seen in Fig.\ref{fig:renyi-n}, the Hartley entropy is divergent in this case. On the other hand in $n\rightarrow \infty$ limit one finds the min-entropy
\begin{align}
S^{(\infty)}=\lim_{n\rightarrow \infty} S^{(n)}=-\log \lambda_{\rm{max}},
\end{align}
where $\lambda_{\rm{max}}$ is the largest eigenvalue of the reduced density matrix. In this case according to Fig.\ref{fig:renyi-n} the Renyi entropy saturates to a constant value which depends on the value of the coupling $\lambda$, as expected. Also note that in all cases the Renyi entropy is a decreasing function of the Renyi index $n$. 
\subsection{Entanglement Inequalities}
In a general quantum-mechanical system or quantum field theory, entanglement entropy (and other measures of quantum entanglement) are proved to satisfy various inequalities. As a first example of such inequalities, we consider those dealing with Renyi entropy which was defined in Eq.\eqref{eq:EERE}. Renyi entropies must satisfy a variety of different inequalities such as \cite{Beck}
\begin{align}\label{eq:renyiinequ}
&\frac{\partial}{\partial n} S^{(n)}\leq 0, \hspace*{2.2cm}\frac{\partial }{\partial n}\left((n-1)S^{(n)}\right)\geq 0,\nonumber\\
&\frac{\partial }{\partial n}\left(\frac{n-1}{n}S^{(n)}\right)\geq 0, \hspace*{0.5cm}\frac{\partial^2 }{\partial n^2}\left((n-1)S^{(n)}\right)\leq 0.
\end{align}
\begin{figure}
\begin{center}
\includegraphics[scale=.7]{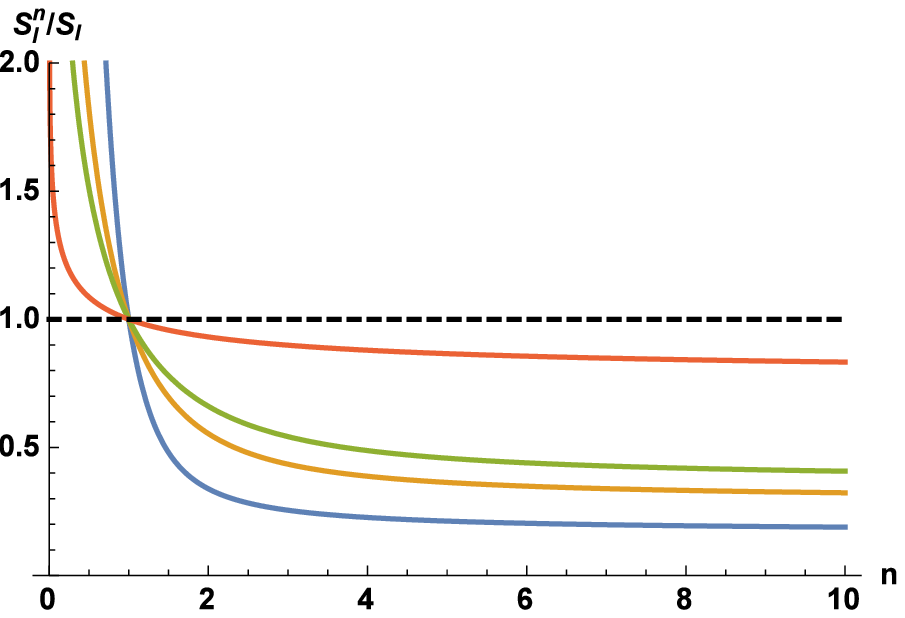}
\includegraphics[scale=.7]{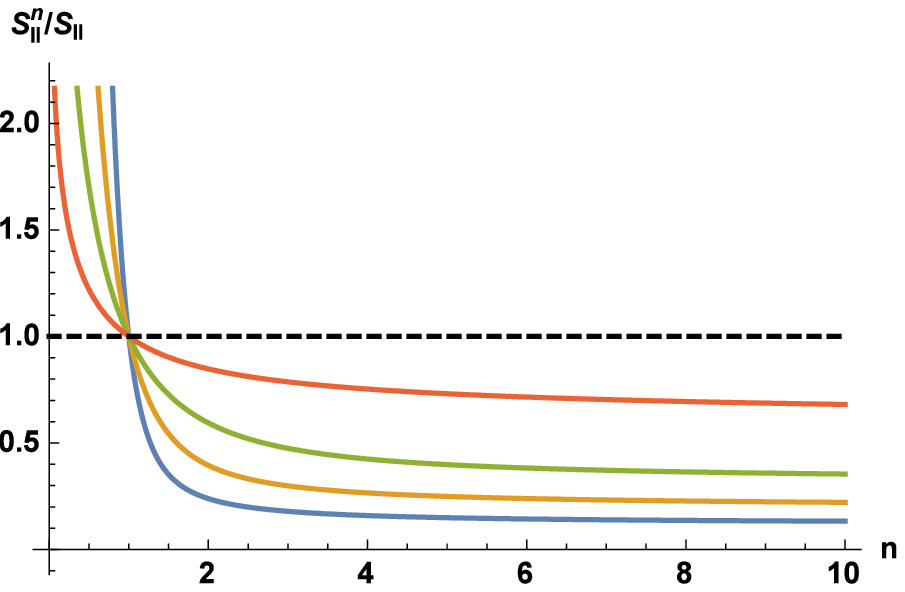}
\end{center}
\caption{Renyi entropy of infinite-range (left) and nearest-neighbour (right) models as a function of Renyi index $n$ for $N=8$ and $m=4$. The dashed black curve corresponds to the value of entanglement entropy.}
\label{fig:renyi-n}
\end{figure}
As we mentioned before, the first inequality which shows Renyi entropy is a decreasing function of Renyi index $n$ is satisfied in our models (see Fig.\ref{fig:renyi-n}). It is a straight forward exercise to show that the other three inequalities are also satisfied in both of our models. 

In what follows in this subsection we consider other important inequalities which is expected to be satisfied generally, based on the classification given in \cite{Headrick:2013zda}: 

\vspace{3mm}
1) $S_A\geq 0$ (positivity of EE)

This is a trivial property which we have checked it for different points in the parameter space of our models in the previous section (see Fig.\ref{fig:EEMIlambda} and Fig.\ref{fig:EEMIIlambda}).

\vspace{3mm}
2) $S_A+S_B\geq S_{A\cup B}$ (Subadditivity)

This property can be rephrased in terms of the positivity of mutual information (MI) which is defined as\footnote{Note that the definition of MI does not restrict subsystems $A$ and $B$ to be complements.}
\be
I(A,B)=S_A+S_B-S_{A\cup B}.
\ee
MI is a quantity which measures the amount of shared information between $A$ and $B$. While dealing with SEE, where $A$ and $B$ correspond to spatial subregions, MI is a UV finite measure of entanglement in contrast to EE. Clearly the subadditivity property implies the positivity of MI, i.e., $I(A,B)\geq 0$. Using the definition of Renyi entropy, one can also define mutual Renyi information (MRI) from the corresponding Renyi entropies as
\be
I^{(n)}(A,B)=S^{(n)}_A+S^{(n)}_B-S^{(n)}_{A\cup B}.
\ee
While dealing with SEE, it is known that MRI does not have a definite sign. It might be interesting to verify this property in the case of FSEE.
\begin{figure}
\begin{center}
\includegraphics[scale=.75]{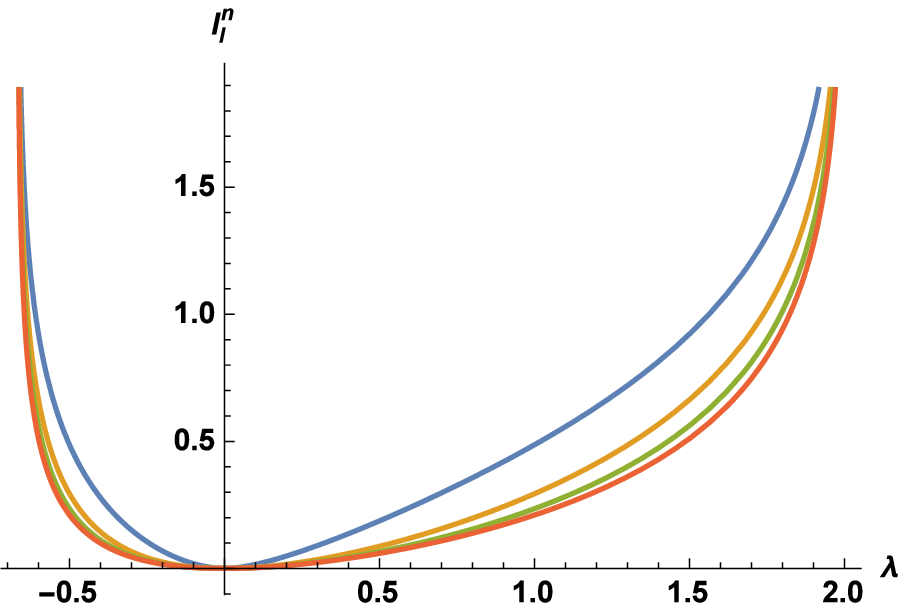}
\includegraphics[scale=.75]{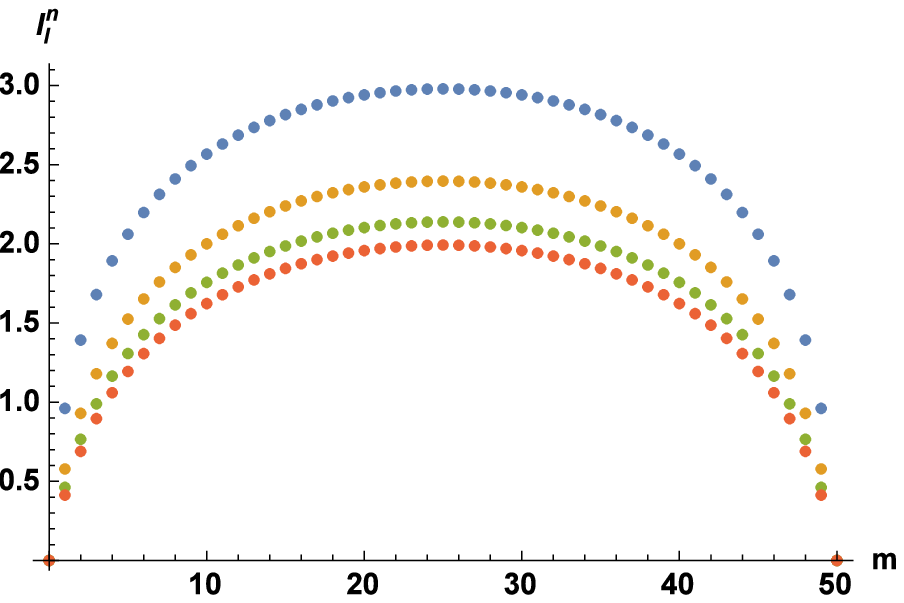}
\end{center}
\caption{Mutual Renyi information of infinite-range model. \textit{Left}: MRI as a function of coupling $\lambda$ for $N=4,m_1=1, m_2=2$ and different value for Renyi index. \textit{Right}: MRI as a function of $m$ for $\lambda=0.9$ and $N=50,m_1=m, m_2=50-m$ with the same value of Renyi index.}
\label{fig:mutual-I}
\end{figure}
\begin{figure}
\begin{center}
\includegraphics[scale=.75]{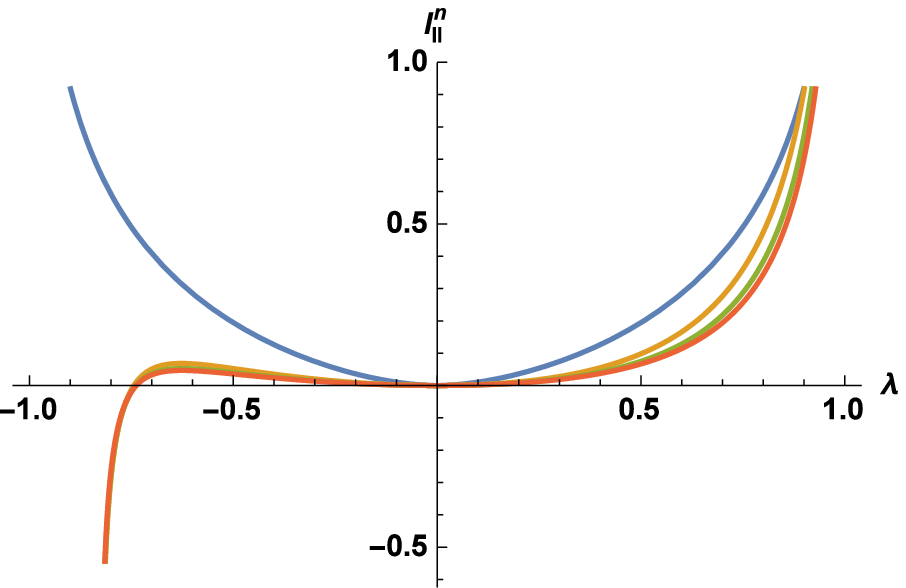}
\includegraphics[scale=.75]{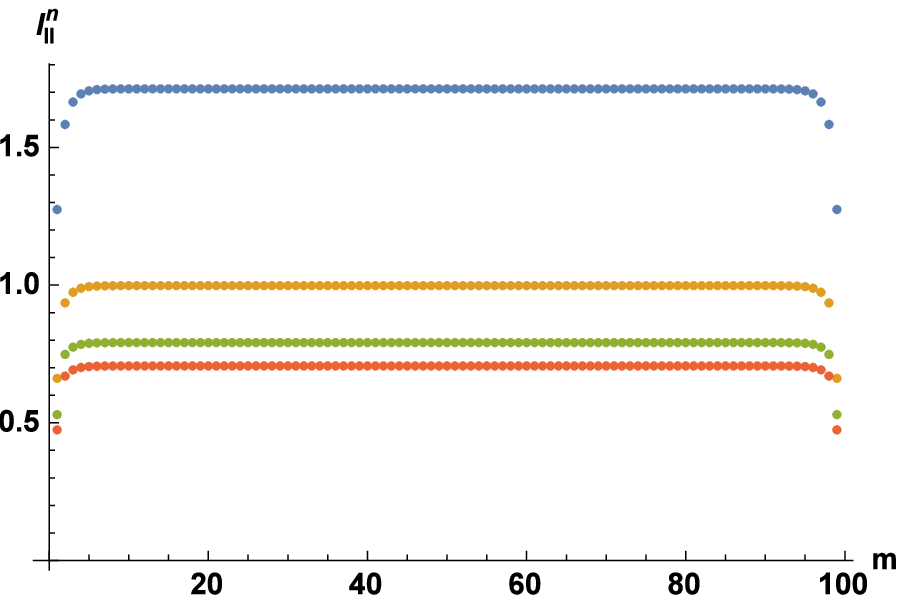}
\end{center}
\caption{Mutual Renyi information of nearest-neighbour model. \textit{Left}: MRI as a function of coupling $\lambda$ for $N=4,m_1=1, m_2=2$ and different value for Renyi index. \textit{Right}: MRI as a function of $m$ for $\lambda=0.9$ and $N=50,m_1=m, m_2=50-m$ with the same value of Renyi index. }
\label{fig:mutual-II}
\end{figure}

Since we are dealing with FSEE, the Hilbert space decomposition we chose implied $I(m_1,m_2)=S_{m_1}+S_{m_2}-S_{m_1+m_2}$, where $S_{m_i}$ is the FSEE for the case which we have integrated out $(N-m_i)$ fields (similarly for MRI). We have plotted MI and MRI for both the infinite-range and the nearest-neighbour models in Fig.\ref{fig:mutual-I} and Fig.\ref{fig:mutual-II} where we have considered the $\lambda$ and $m$-dependence of these quantities. In both of these figures, the blue curve corresponds to the case of MI, and other curves correspond to higher Renyi indices, i.e. MRI. MI is shown to be always positive in our models. It is worth to note that we could not find any region in the parameter space of the infinite-range model where the MRI admits negative values. The typical behavior of this quantity is similar to what is shown in Fig.\ref{fig:mutual-I} for specific values of the parameters. In the nearest-neighbour model the MRI have both positive and negative values as shown in Fig.\ref{fig:mutual-II}. Note that while we deal with $m_1$ and $m_2$ which are complements, we expect the MRI to be symmetric with respect to half of the whole number of fields denoted by $N$ (see the right plots in Fig.\ref{fig:mutual-I} and Fig.\ref{fig:mutual-II}). 
\begin{figure}
\begin{center}
\includegraphics[scale=.7]{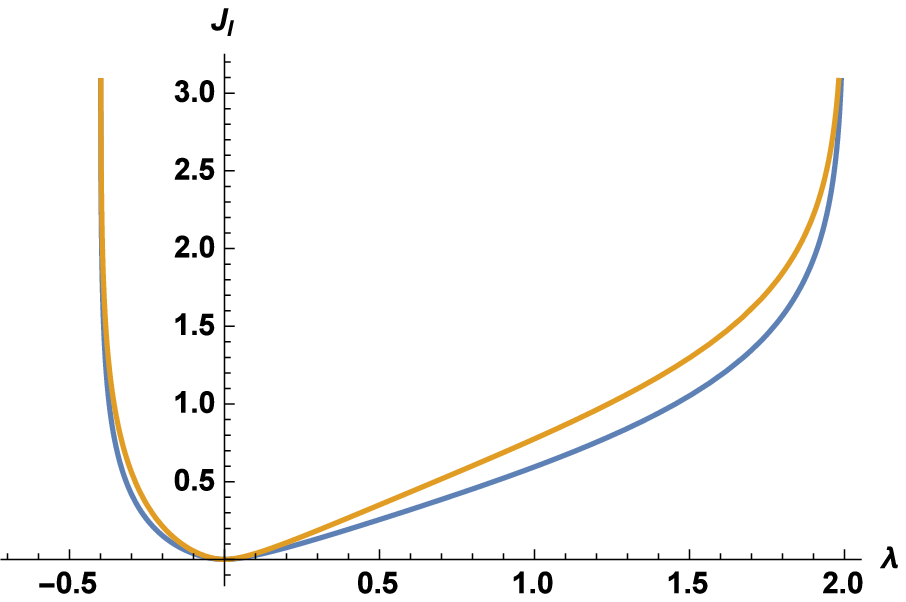}
\includegraphics[scale=.7]{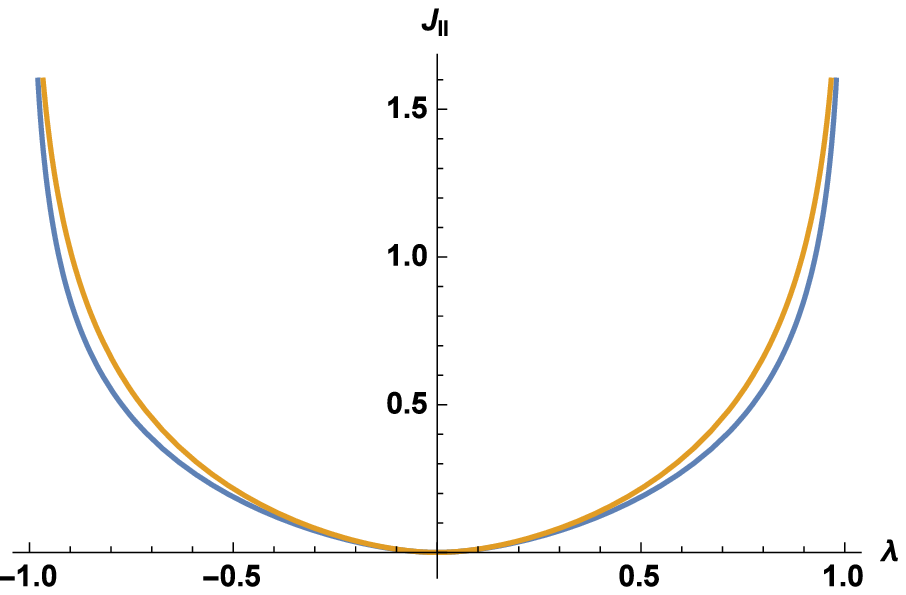}
\end{center}
\caption{Intrinsic entropy of infinite-range and nearest-neighbour models as a function of coupling $\lambda$ for $N=6,m_1=1$ and different values of $m_2$.}
\label{fig:intrinsic}
\end{figure}

\vspace{3mm}
3) $S_A\leq S_{A\cup B}+S_B$ (Araki-Lieb inequality)

This property which is also called the triangle inequality implies the positivity of the intrinsic entropy which is defined as
\be
J(B,A)=S_{A\cup B}+S_B-S_A\;\;\;\; , \;\;\;\;J(B,A)\geq 0.
\ee
Some specific examples of this inequality in our models are depicted in Fig.\ref{fig:intrinsic}.

\vspace{3mm}
4) $S_{A\cup B\cup C}+S_B\leq S_{A\cup B}+S_{B\cup C}\;\;\;\;,\;\;\;\;S_A+S_C\leq S_{A\cup B}+S_{B\cup C}$ (Strong subadditivity)

\begin{figure}
\begin{center}
\includegraphics[scale=.75]{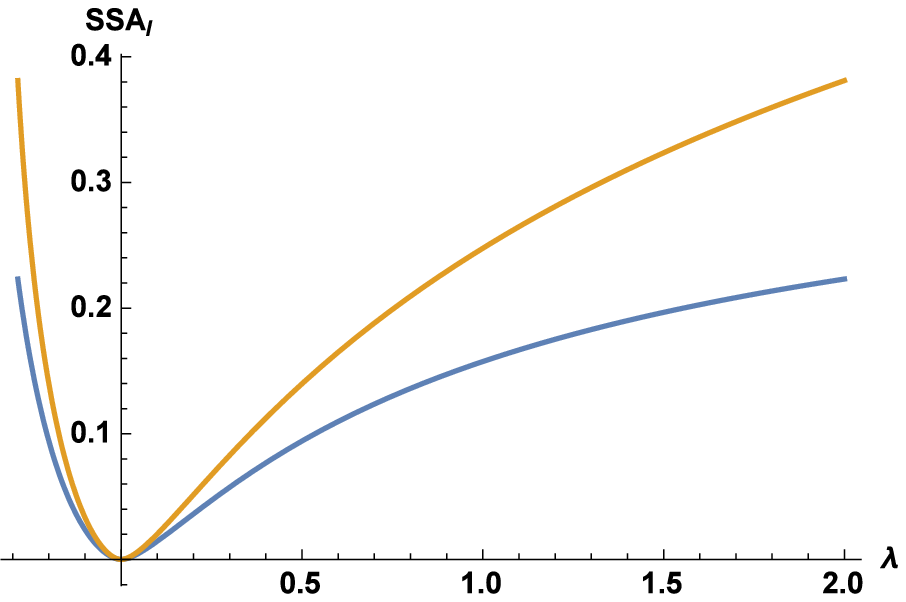}
\includegraphics[scale=.75]{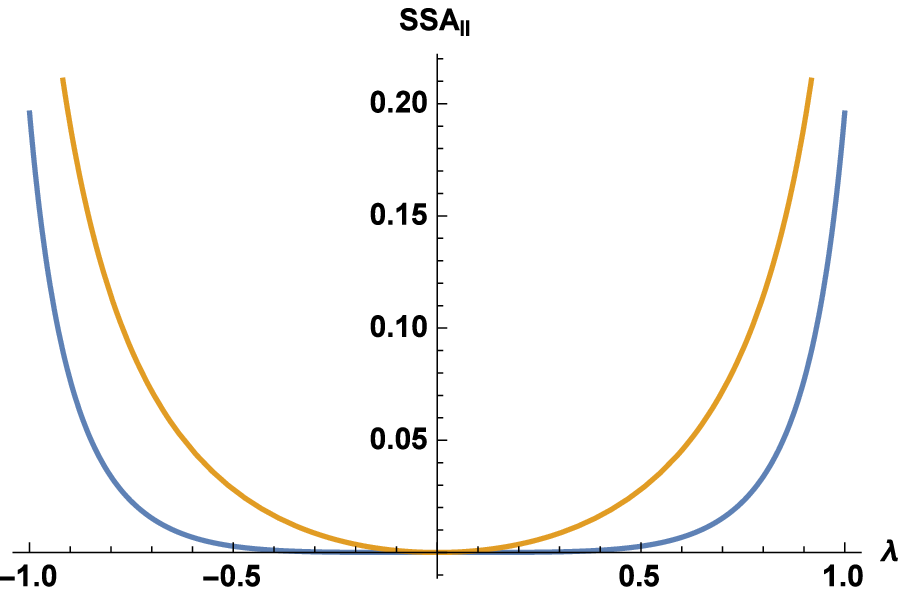}
\end{center}
\caption{SSA inequalities in infinite-range model (left) and nearest-neighbour model (right) as a function of coupling $\lambda$ for $N=8,m_1=1, m_2=2, m_3=3$.}
\label{fig:SSA}
\end{figure}

Both of these inequalities are called strong subadditivity (SSA) and must hold in any quantum system. These inequalities physically mean that mutual information and intrinsic entropy must increases under inclusion. These inequalities hold in our models as we have plotted explicit examples of them in both of our models in Fig.\ref{fig:SSA}.

\vspace{3mm}
5) $S_A+S_B+S_C+S_{A\cup B\cup C}\leq S_{A\cup B}+S_{A\cup C}+S_{B\cup C}$ (Monogamy of mutual information (MMI))

\begin{figure}
\begin{center}
\includegraphics[scale=.7]{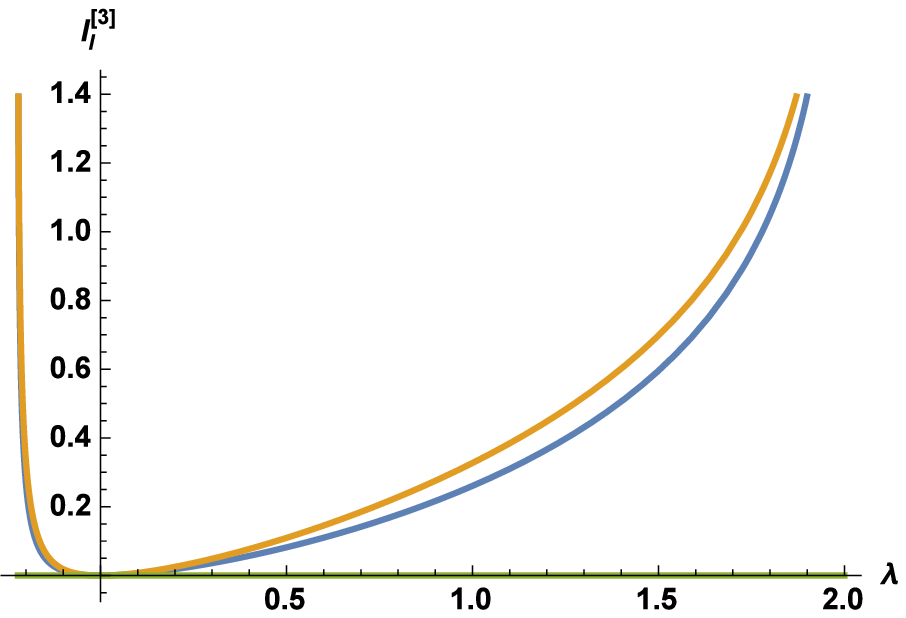}
\includegraphics[scale=.7]{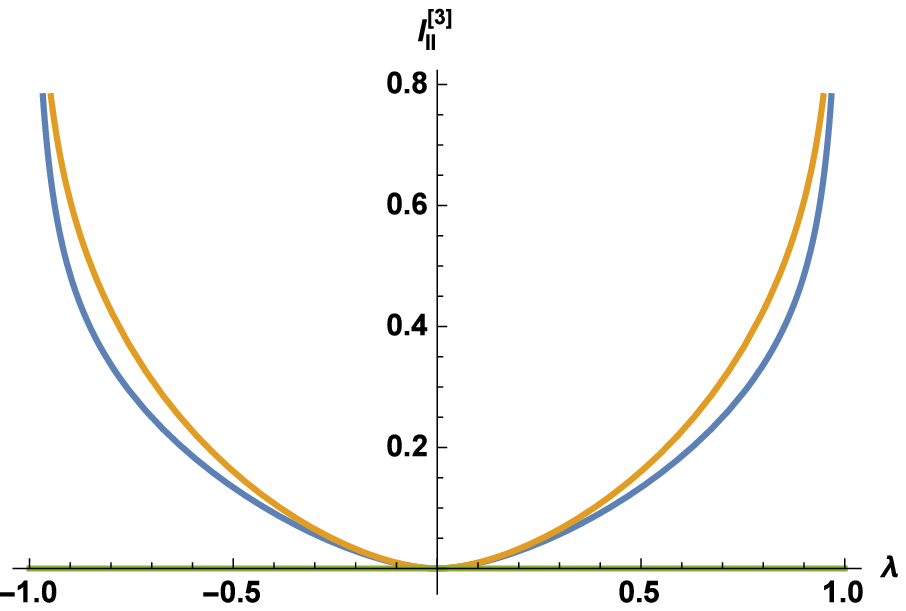}
\end{center}
\caption{Tripartite information for infinite-range and nearest-neighbour models as a function of coupling $\lambda$ for $N=10,m_1=1, m_2=3$ and different values of $m_3$. Note that $I^{[3]}$ is always positive and in the latter case saturates to zero.}
\label{fig:tripartite-I}
\end{figure}

In spite of previously mentioned inequalities, which are general properties of entanglement measures in any quantum system, MMI does not necessarily hold in any quantum system and thus it is not considered as feature of entanglement entropy. Again this inequality can be rephrased as the negativity of tripartite information, i.e. $I^{[3]}(A, B, C)\leq 0$, which is defined as 
\begin{align}\label{eq:3partite}
I^{[3]}(A, B, C)&=S_A+S_B+S_C-S_{A\cup B}-S_{A\cup C}-S_{B\cup C}+S_{A\cup B\cup C}\\\nonumber
&=I(A, B)+I(A, C)-I(A, B\cup C).
\end{align}
Generally in quantum mechanics or even in QFTs, depending on how the Hilbert space is partitioned, $I^{[3]}$ can be positive, negative or zero. In Fig.\ref{fig:tripartite-I} we have plotted $I^{[3]}$ for both of our models corresponding to different partitioning of the field space. As is shown in Fig.\ref{fig:tripartite-I}, this inequality does not hold in both of our models and more interestingly the tripartite information is always non-negative in these models. It is also interesting to note that in the case of  $m_1+m_2+m_3=N$ the tripartite information becomes zero. According to second equality of \eqref{eq:3partite} this is a reminiscent of models which exhibit extensive mutual information property \cite{Casini:2008wt}.

\subsection{$n$-partite Information}
In the context of quantum information theory, partitioning the system into $n$-parts, a new quantity known as $n$-partite information\footnote{This $n$ has nothing to do with the index of Renyi entropy.}
is defined as \cite{Hayden:2011ag}
\begin{align}\label{eq:npartite}
I^{[n]}(A_{\{i\}})=\sum_{i=1}^nS_{A_i}-\sum_{i<j}^n S_{A_i\cup A_j}+\sum_{i<j<k}^n S_{A_i\cup A_j\cup A_k}
-\cdots\cdots +(-1)^n S_{A_1\cup A_2\cup\cdots\cup A_n}.
\end{align}
It is obvious that according to the above formula the definition of 1-partite and 2-partite information reduce to EE and MI respectively. Also note that the $n$-partite information for $n>1$ is a UV finite quantity. Actually a finite measure for quantum entanglement between subsystems of a larger system is not unique (e.g. another choice known as multi-partite information is defined in \cite{Horodecki:2009zz}). The reason why we use the above definition for $n$-partite information Eq.\eqref{eq:npartite} for such a quantity is due to its property which reduces to the definition of tripartite information Eq.\eqref{eq:3partite} in the case of $n=3$ (while e.g. multi-partite information does not have this property \cite{Horodecki:2009zz}). 

As we have mentioned before, MI is always non-negative, i.e., $I^{[2]}\geq0$, due to the subadditivity property of EE. Although the sign of tripartite information is not fixed generally, but as we have shown in the previous subsection it is always non-negative in both of our models. It is worth to note that in the case of CFTs which support a gravitational dual, the sign of tripartite information is fixed to be always negative. This general property restricts the holographic mutual information to be monogamous \cite{Hayden:2011ag}.\footnote{It is also shown in reference \cite{Allais:2011ys} that the null energy condition is a necessary condition for the monogamy of holographic mutual information.} As an extension of this property, it is also shown in reference \cite{Alishahiha:2014jxa} that in a specific limit in the case of SEE, the holographic $n$-partite information has a definite sign: it is positive (negative) for even (odd) $n$.

\begin{figure}
\begin{center}
\includegraphics[scale=.7]{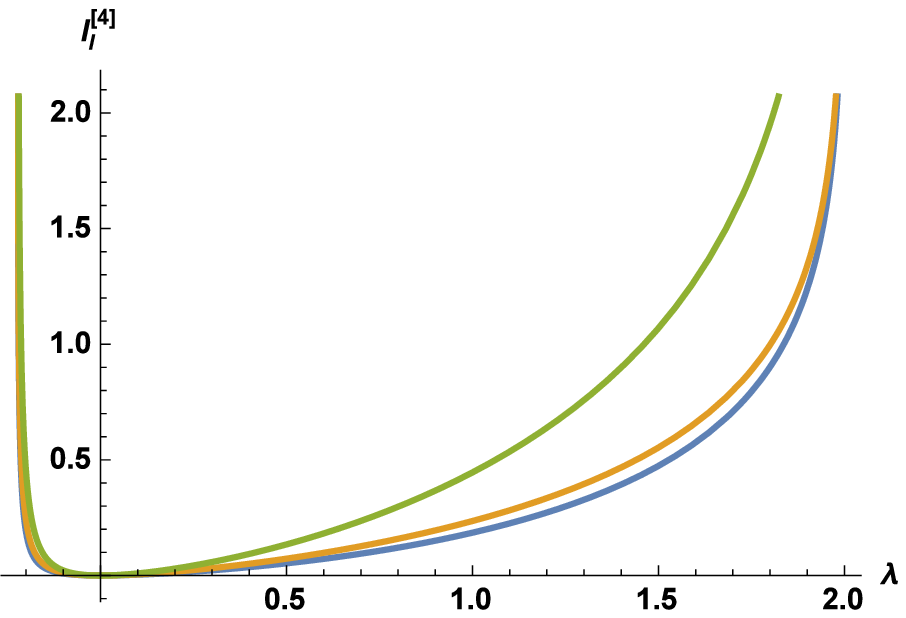}
\includegraphics[scale=.7]{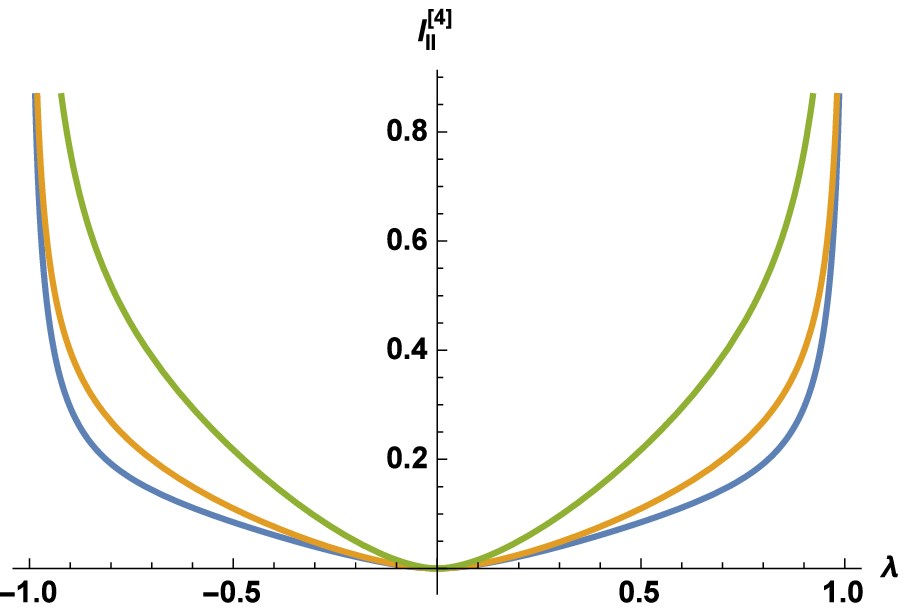}
\end{center}
\caption{4-partite information for model I and II as a function of coupling $\lambda$ for $N=10, m_1=m_2=m_3=1$ and different values of $m_4$.}
\label{fig:i4}
\end{figure}

\begin{figure}
\begin{center}
\includegraphics[scale=.7]{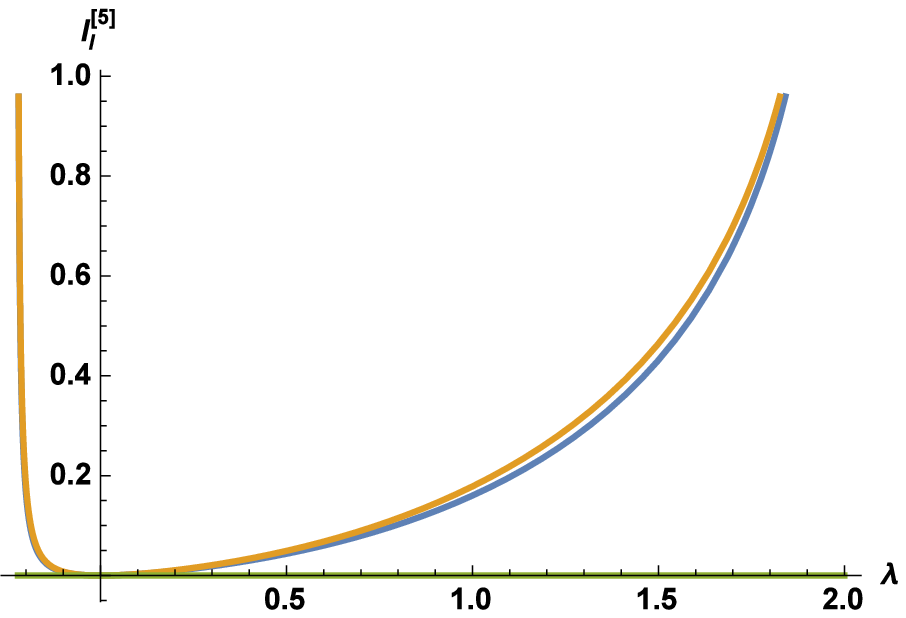}
\includegraphics[scale=.7]{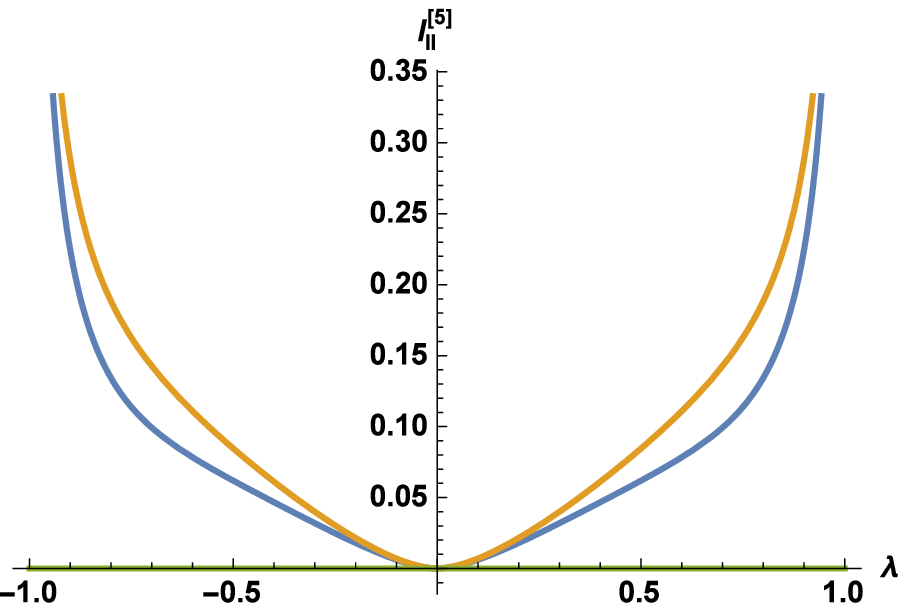}
\end{center}
\caption{5-partite information for model I and II as a function of coupling $\lambda$ for $N=10, m_1=m_2=m_3=m_4=1$ and different values of $m_5$. Note that $I^{[5]}$ is always positive and in the latter case saturates to zero.}
\label{fig:i5}
\end{figure}

It would be interesting to investigate the sign of higher $n$-partite information in our models. In Fig.\ref{fig:i4} and Fig.\ref{fig:i5} we present the 4-patite and 5-partite information as a function of the coupling $\lambda$ which is surprisingly always positive. Also focusing on 5-partite information together with 3-partite information (see Fig.\ref{fig:tripartite-I}), one may conjecture that $n$-partite information is always vanishing for the case of odd $n$'s with complement partitioning of the system i.e. $\sum_i m_i=N$. 
\subsection{Entanglement Negativity}
Entanglement negativity and its counterpart logarithmic negativity are useful measures of quantum entanglement even for mixed states \cite{Vidal:2002}. It is known that the von-Neumann entropy for a mixed state, e.g. a thermal state, dominated by the classical correlations is not a useful measure for quantum entanglement. MI also measures the total correlations (both quantum and classical) between two subsystems which just offers an upper bound \cite{Wolf:2008}. It has been shown that negativity is an  entanglement monotone (does not increase under any LOCC operations) and hence a proper measure for quantum entanglement \cite{Plenio:2005}.
To give a more concrete but nevertheless simple definition of this quantity one may consider a tripartite system in a pure state with a complement partitioning, i.e., $M=A_1\cup A_2\cup A_3$. In this case the reduced density matrix corresponding to union of two subsystems is described by a mixed state $\rho\equiv \rho_{A_1\cup A_2}$. Entanglement negativity and logarithmic negativity are defined as
\begin{align}\label{negativity}
\mathcal{N}(\rho)\equiv \frac{\Vert \rho^{T_2}\Vert-1}{2},\;\;\;\;\;\mathcal{E}(\rho)=\log \Vert \rho^{T_2}\Vert,
\end{align}
where $\Vert \rho^{T_2}\Vert$ denotes the trace norm of the partial transpose of $\rho$. With the above definition the logarithmic negativity measures how much the eigenvalues of $\Vert \rho^{T_2}\Vert$ are negative.

\begin{figure}
\begin{center}
\includegraphics[scale=.7]{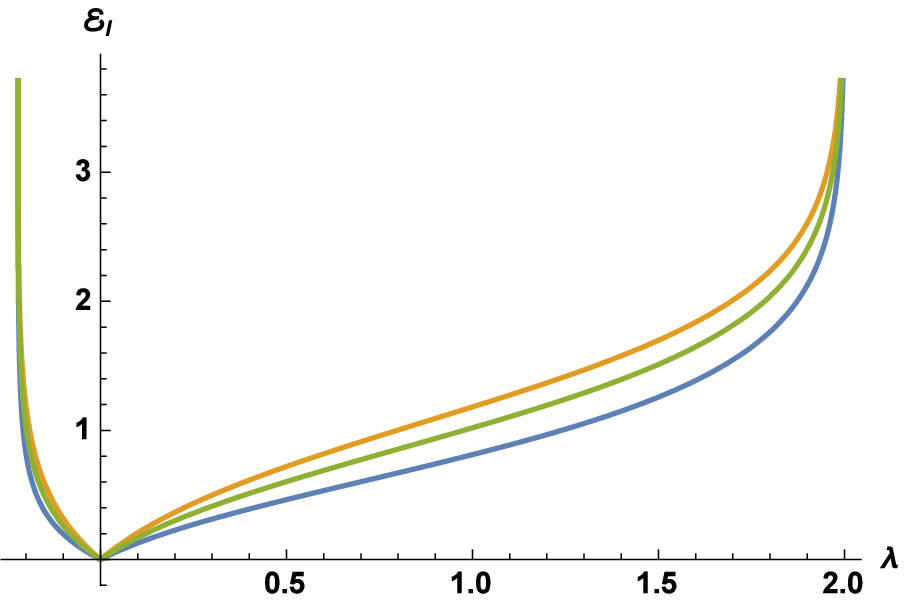}
\includegraphics[scale=.7]{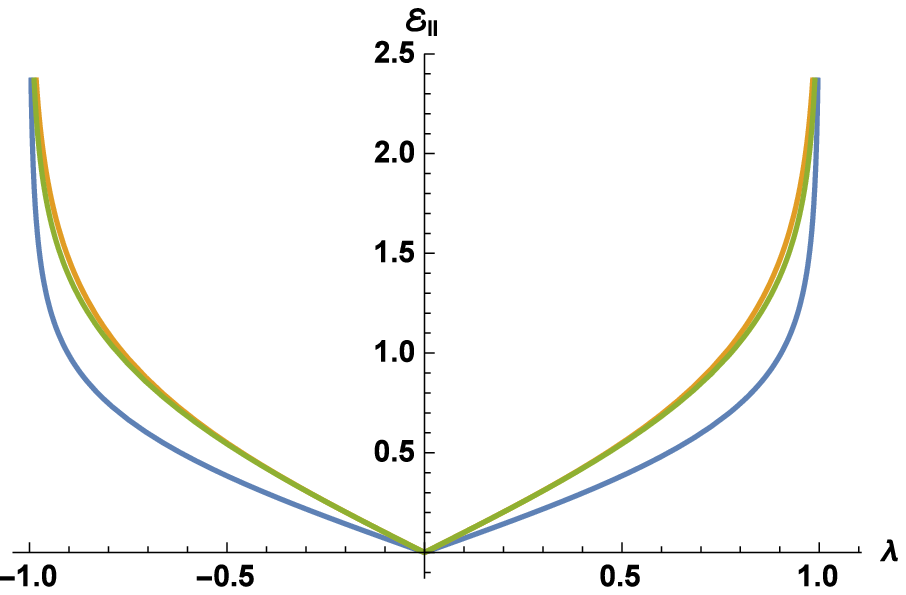}
\end{center}
\caption{Logarithmic negativity for infinite-range and nearest-neighbour models as a function of coupling $\lambda$ for $N=10$ and different values of $m$.}
\label{fig:negativity}
\end{figure}

Although computing these quantities in general is not an easy task, the authors of \cite{Calabrese:2012ew} have introduced a replica approach to obtain the logarithmic negativity in the ground state of 2d CFTs. They also show that for a pure state and bipartite system where $\mathcal{H}=\mathcal{H}_1\otimes \mathcal{H}_2$, this quantity is given by Renyi entropy with $n=1/2$, i.e., 
\begin{align}\label{negativityrenyi}
\mathcal{E}(\rho_2)=2\log \Tr \rho_2^{1/2}.
\end{align}
We focus on this definition in order to study the logarithmic negativity in our models. We postpone further investigations based on computing Eq.\eqref{fig:negativity} for future works. In Fig.\ref{fig:negativity} we have plotted logarithmic negativity as a function of coupling $\lambda$ for different partitions of the Hilbert space. 

\section{Conclusions and Discussions}\label{sec:dis}
In this paper we have considered a less studied type of entanglement which is known as field space entanglement. This type of entanglement corresponds to a Hilbert space decomposition in the field space of a quantum field theory. As a simple laboratory to study field space entanglement, we have considered a theory with a generic $N$ number of free scalar fields, we added kinetic mixing terms (in terms of two specific models) which generates entanglement between these scalar fields. We traced out a generic $m$ number of these fields and worked out the entanglement and Renyi entropies between $m$ and $(N-m)$ number of these scalar fields. The result of these entropies is UV-divergent which scales with the (spatial) volume of the theory as expected. Similar to the case of spatial entanglement entropy, there is a universal term, i.e. a UV cut-off independent term which we argue to carry some information about the theory. Beside the entanglement and Renyi entropies, we also constructed other well known entanglement measures such as mutual information, intrinsic entropy and $n$-partite information to further investigate features of field space entanglement. We have shown that this type of entanglement in our models satisfy most of the known general features of entanglement measures including Renyi entropy inequalities, strong subadditivity and Araki-Lieb inequality. We have also studied the monogamy of mutual information which has a definite sign (positive) for tripartite, 4-partite, and 5-partite information in our models.

There are several directions which one can follow to further investigate our models and the notion of field space entanglement using this laboratory. We leave further investigations of these models, including the recently proposed entanglement inequalities (see \cite{Bao:2015bfa}), to future works and in the following of this section we discuss a few words about the holographic picture of field space entanglement entropy and also offer a different viewpoint to this family of field theories which we have considered.
\subsection*{Holographic Picture of FSEE}
In order to gain some information about the possible gravity picture of such an analysis, as the first step we consider some well known features of field theories which support holographic duals: the monogamy condition for holographic mutual information and its implication on the dual field theory. As we mentioned in the previous section the tripartite information in both of our models is always positive and the monogamy constraint does not hold. Actually this behavior is in contrast to the holographic result which shows that the holographic mutual information is always monogamous \cite{Hayden:2011ag}. So in this sense it seems that our models do not have a well defined holographic description. It is important to mention that it is not clear that whether this constraint must hold for any type of EE or it is just a feature of SEE. In the following for a while we forget about this comment on the relation between monogamy of mutual information and the existence of a holographic dual.

The authors of reference \cite{Mollabashi:2014qfa} have proposed a naive holographic picture for the entanglement entropy between two CFTs which might be related to our models in the case of $N=2$. In this proposal the factorization of the Hilbert space in the field space was related to partitioning the compact part of the AdS$_5\times$S$^5$ geometry by introducing a $\p A$ surface which partitions the S$^5$ sphere into two parts and wraps the boundary of AdS$_5$. The minimal surface anchoring the corresponding boundary on a certain UV cut-off surface was proposed to give the entanglement entropy between two interacting subsectors of the whole CFT$_4$ (which is dual to the AdS$_5\times$S$^5$ geometry). Although there are some substantive comments about the relation between this holographic picture and FSEE (see \cite{Taylor:2015kda} and also \cite{Karch:2014pma}), the holographic dual of our models in this picture is straightforward. One may partition the S$^5$ sphere to $N$ parts and the corresponding entanglement entropy is proportional to the volume of different portions. For example if we consider the mutual information between two set of fields, the S$^5$ sphere is divided into three parts and different terms contributing in the expression of mutual information are proportional to the volume of the corresponding part of the sphere.

There is another geometrical picture introduced in reference \cite{Karch:2014pma} which offers a geometrical interpretation for the entanglement between two SU($m$) and SU($N-m$) CFTs again as subsectors of the dual CFT$_4$. This picture is based on the interpretation of minimal surfaces in the more general supergravity Coulomb branch geometry rather than AdS$_5\times$S$^5$ as entanglement entropies. Here the level sets of the scale factor multiplying the Minkowski part of the solution is interpreted as the UV cut-off of the CFTs living on separated stacks of D3-branes. There are two family of level sets: disconnected level sets which are consisted of two separated surfaces surrounding each brane stack, and connected ones which are single surfaces surrounding both brane stacks. Correspondingly there are two family of minimal surfaces, those which start and end on the connected level sets and those which start and end on the disconnected level sets. Those surfaces which start and end on the connected level sets are interpreted as a  measure for the entanglement between two CFTs living on the brane stacks which is generated by means of the stretched modes between these stacks. The minimal surfaces starting and ending on a part of the disconnected level set around, say stack 1, are interpreted as a measure for the entanglement between a part of CFT$_1$ and CFT$_2$ living on the other stack together with the entanglement between two parts of CFT$_1$. For more details see reference \cite{Karch:2014pma}.

One can naively generalize this picture to be appropriate for interpreting mutual information between any two of three SU($m_1$) and SU($m_2$) and SU($N-m_1-m_2$) CFTs by considering three stacks of D3-branes. In this case the number of connected and disconnected level sets increase. There are four types of disconnected level sets: a single one composed of three parts and those which are composed of two parts, one surrounding two stacks and the other surrounding a single stack. Although this configuration for three stacks is too complicated to calculate, there are several minimal surfaces which could be interpreted as a direct generalization of what was discussed in the previous paragraph. One can in principle even generalize this picture for arbitrary $N$ and interpret the corresponding minimal surfaces as in the case of $N=2$ as a possible holographic picture of our models.

On the other hand it is recently argued in reference \cite{Taylor:2015kda} that it is not possible to give a precise geometrical realization for FSEE in a holographic dual and all which is discussed in the above two scenarios is rather related to entanglement in the space of the global symmetry of the CFTs which is in no way essential to define FSEE. Although the author has offered some arguments to give an effective realization to such a case in terms of IR CFTs as dual field theories for internal throats in the Coulomb branch supergravity solution of separated D3-branes, the geometrical interpretation for FSEE seems to still be an open problem.

Now lets forget about different scenarios as candidates for the holographic picture of FSEE. One may focus on the $N$-dependence of the entanglement entropy in the infinite-range model to give a concrete expectation for a possible reliable holographic dual.\footnote{We thank Shahin Sheikh-Jabbari because of his valuable comment about the $N$-dependence of field space entanglement entropy which was insightful for us to clarify the structure of our analysis.} To avoid unnecessary complications, we consider the entanglement entropy in the leading order of $\lambda$
\be
S(m)=\frac{\lambda^2}{32}m(N-m)\left[1-\log\frac{\lambda^2 m(N-m)}{32}\right]+\mathcal{O}\left(\lambda^3\right),
\ee 
which for the special case of $m=\frac{N}{2}$ gives
\be
S(m)=\frac{\lambda^2N^2}{128}\left[1-\log\frac{\lambda^2N^2}{128}\right]+\mathcal{O}\left(\lambda^3\right),
\ee 
which is expected to be explained by any holographic dual. One can work out the corresponding expressions for the nearest-neighbour model.

Beside this check, the large $N$ behavior of these models seems to have interesting features in the field space. In this limit the infinite-range model seems to behave as a non-local theory in the field space while the nearest-neighbor model resembles a local theory.\footnote{We thank Shahin Sheikh-Jabbri for drawing our attention to this interesting point.}  It would be interesting to investigate this property more precisely and study its implications specifically on entanglement and Renyi entropies.
\subsection*{A Model for Black-hole Radiation}
A field theory which consists of a number of interacting fields could be a field theoretic counterpart of Page's model for black-hole evaporation process \cite{Page:1993wv}.\footnote{We thank Mohsen Alishahiha for bringing our attention to this point.} A first and simple clue for this argument is the symmetric behaviour of the entanglement entropy around $m=\frac{N}{2}$ (see Fig.\ref{fig:EEMI} were we have plotted this behavior for both of our models) and one may compare it with the entanglement (or information) evolution during the black-hole evaporation.

In reference \cite{Page:1993wv} the author has considered two subsystems with Hilbert space dimensions $m$ and $n$ respectively such that the total Hilbert space with dimension $m\times n$ is in a pure state. He has shown that the entanglement entropy between these two subsystems is symmetric as a function of the thermodynamical entropy which is defined by $\log m$. Another important result of such a consideration is that the deviation of the entanglement entropy from its maximum value (the thermodynamical entropy), which is defined as ``information", remains almost zero until the entanglement entropy reaches its maximum value.

\begin{figure}
\begin{center}
\includegraphics[scale=.7]{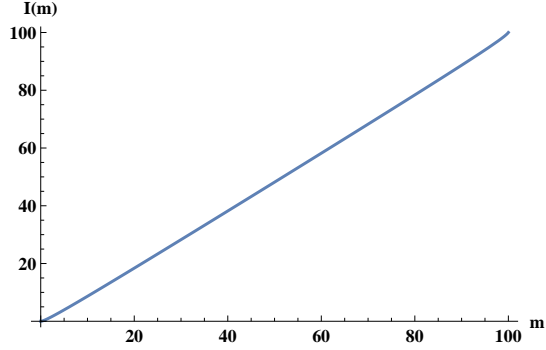}
\end{center}
\caption{The ``information" for the infinte-range model for $N=100$ and $\lambda=0.9$.}
\label{fig:EEMIC}
\end{figure}

We demonstrate the entanglement entropy (see Fig.\ref{fig:EEMI}) and ``information" (see Fig.\ref{fig:EEMIC}) as a function of $m$. The information is defines as $I=m-S$. Our argument for considering such a definition for information in this case is as follows:
In our model where the total Hilbert space includes $N$ fields, the subsystems (I) and (II) have $m$ and $(N-m)$ fields respectively and the thermodynamical entropy is an extensive  quantity. To see this consider the Hilbert space for the first subsystem which is $\mathcal{H}_{(I)}=\mathcal{H}_1 \otimes \mathcal{H}_2\otimes \cdots \otimes \mathcal{H}_m$, so if we denote the dimension of the Hilbert space for a single field by $D$, then the dimension of $\mathcal{H}_{(I)}$ becomes $D^m$. So in our case the themodynamical entropy becomes $\log D^m=m\log D$ and we expect that in the definition of information one must replace $\log m$ with $m$.

Note that in Fig.\ref{fig:EEMIC} which we have plotted the information $I$, it is non-zero even in the early stages of evolution $(m\sim 1)$, in contrast with what was previously found in \cite{Page:1993wv}.
\subsection*{Acknowledgements}
It is our pleasure to thank Mohsen Alishahiha, Joey Medved, Shahin Sheikh-Jabbari, Noburo Shiba, Tadashi Takayanagi and Cenke Xu for valuable discussions and correspondence. We also thank John Cardy, Saman Moghimi-Araghi, Ali Naji and Shinsei Ryu for their correspondence about possible relations between our models and statistical physics. We also thank Mohsen Alishahiha, Mukund Rangamani and Shahin Sheikh-Jabbari for careful reading of the manuscript and their comments on the draft.
The authors thank the organizers of ``IPM String School and Workshop 2015" where the early stages of this work took place and also Kavli IPMU and ISSP for organizing ``International Workshop on Condensed Matter Physics and AdS/CFT" at university of Tokyo for warm hospitality during parts of this work. This work is supported by Iran National Science Foundation (INSF).  

\appendix
\section{Calculation of Reduced Density Matrix}\label{sec:app1}
In this section we explain some details of the calculation of our master formula, which is the trace of the reduced density matrix of both of our models reported in Eq.\eqref{eq:TrrhonI} and Eq.\eqref{eq:TrrhonII}. Here we explain the logical steps with general formulas as the key points leading to these results. The remaining part, although is some how messy, it is of course straightforward if one follows the procedure discussed in this section. The starting point is the wave functional for Gaussian models introduced in Eq.\eqref{eq:WF}. We explain the general formalism while explaining the infinite-range model in subsection \ref{sec:AIR}, and turn to the nearest-neighbour model in subsection \ref{sec:ANN}. 
\subsection{Infinite-Range Model}\label{sec:AIR}
As we have mentioned in Sec.\ref{sec:KMGM}, the total density matrix of these models is generally defined as
\be\label{eq:rhomA}
\rho_{\mathrm{tot.}}[\phi'_{1},\phi_{1},\phi'_{2},\phi_{2},\cdots,\phi'_{N},\phi_{N}]=\Psi^*[\phi'_1,\phi'_2,\cdots,\phi'_N]\Psi[\phi_1,\phi_2,\cdots,\phi_N],
\ee
where $\Psi[\{\phi\}]$ is the Gaussian wave functional introduced in Eq.\eqref{eq:WF}.
In order to define the reduced density matrix for the simplest case, i.e. $m=1$, we identify $\phi_1$ and $\phi'_1$ and integrate over it on the whole space
\be\label{eq:rhomA2}
\rho_{(N-1)}[\phi'_{2},\phi_{2},\phi'_{3},\phi_{3},\cdots,\phi'_{N},\phi_{N}]=\int\mathcal{D}\phi_1\,\Psi^*[\phi_1,\phi'_2,\cdots,\phi'_N]\Psi[\phi_1,\phi_2,\cdots,\phi_N].
\ee
Implementing the explicit form of the Gaussian wave functional given in Eq.\eqref{eq:WF} and performing the integral, up to an irrelevant normalization constant the result is
\begin{align}\label{eq:rho1A}
\begin{split}
\rho_{(N-1)}[\phi'_{2},\phi_{2}&,\cdots,\phi'_{N},\phi_{N}]=
\exp\Bigg\{-\frac{1}{2}\sum_{i,j=2}^N\int dx^{d-1}dy^{d-1}\Bigg[\phi_i(x)\left(G_{ij}-\frac{G_{1i}G_{1j}}{\widetilde{G_{11}}}\right)\phi_j(y)\\
&+\phi'_i(x)\left(G^*_{ij}-\frac{G^*_{1i}G^*_{1j}}{\widetilde{G_{11}}}\right)\phi'_j(y)
-\phi_i(x)G_{1i}G^*_{1j}\phi'_j(y)-\phi'_i(x)G^*_{1i}G_{1j}\phi_j(y)\Bigg]\Bigg\},
\end{split}
\end{align}
where we have dropped the $x$ and $y$ dependence of $G_{ij}$'s in the above expression for simplicity and we do so in what follows. Note that in the above formula $\tilde{\bullet}\equiv2\mathrm{Re}\left[\bullet\right]$. It is not a hard task to integrate out more than one field, say $m$ number of fields which leads to the reduced density matrix
\be\label{eq:rhomAm}
\rho_{(N-m)}[\phi'_{m+1},\phi_{m+1},\cdots,\phi'_{N},\phi_{N}]=\int\mathcal{D}\phi_1\cdots\mathcal{D}\phi_m\,\Psi^*[\phi_1,\phi'_2,\cdots,\phi'_N]\Psi[\phi_1,\phi_2,\cdots,\phi_N].
\ee
A similar procedure which leads to Eq.\eqref{eq:rho1A} can be performed to arrive at (via induction)
\begin{align}\label{eq:rhomAmm}
\begin{split}
\rho_{(N-m)}[\phi'_{m+1},&\phi_{m+1},\cdots,\phi'_{N},\phi_{N}]=
\exp\Bigg\{-\frac{1}{2}\sum_{i,j=m+1}^N\int dx^{d-1}dy^{d-1}\times\\
&\left[\phi_i(x)X^{(m)}_{ij}\phi_j(y)+\phi'_i(x){X^{(m)}_{ij}}^*\phi'_j(y)
+\phi_i(x){Y^{(m)}_{ij}}\phi'_j(y)+\phi'_i(x){Y^{(m)}_{ij}}^*\phi_j(y)\right]\Bigg\},
\end{split}
\end{align}
where
\be
X^{(m)}_{ij}=X^{(m-1)}_{ij}-\frac{Z^{(m)}_iZ^{(m)}_j}{\widetilde{X^{(m-1)}_{mm}}}\;\;\;\;\;,\;\;\;\;\;
Y^{(m)}_{ij}=Y^{(m-1)}_{ij}-\frac{Z^{(m)}_i{Z^{(m)}_j}^*}{\widetilde{X^{(m-1)}_{mm}}}\;\;\;\;\;,\;\;\;\;\;
Z^{(m)}_i=X^{(m-1)}_{i,m-1}+Y^{(m-1)}_{i,m-1}.
\ee
One can work out the generic reduced density matrix using the above recursion relations with initial values $X^{(0)}_{ij}=G_{ij}$ and $Y^{(0)}_{ij}=0$. Considering the infinite-range model, using Eq.\eqref{eq:rhomAmm} together with Eq.\eqref{eq:Gij1}, one can find the most general form of the reduced density matrix in terms of $m$, $N$ and $\lambda$ which is the coupling constant between the scalar fields. For future use we rewrite the reduced density matrix as
\begin{align}\label{eq:rhomAmm}
\begin{split}
\rho_{(N-m)}[\phi'_{m+1},\phi_{m+1}&,\cdots,\phi'_{N},\phi_{N}]=
\exp\Bigg\{-\frac{1}{2}\int dx^{d-1}dy^{d-1}\times\\
&\begin{pmatrix}\phi'_{m+1}(x)&\phi_{m+1}(x)&\cdots &\phi'_{N}(x)&\phi_{N}(x)\end{pmatrix}\cdot M(m,N) \cdot
\begin{pmatrix}\phi'_{m+1}(y) \\ \phi_{m+1}(y) \\ \vdots \\ \phi'_N(y)  \\ \phi_N(y)\end{pmatrix} \Bigg\},
\end{split}
\end{align}
where
\be
M(m,N)=\begin{pmatrix}
{X^{(m)^*}_{m+1,m+1}} & {Y^{(m)^*}_{m+1,m+1}} &{X^{(m)^*}_{m+1,m+2}} & {Y^{(m)^*}_{m+1,m+2}} & \cdots  &{X^{(m)^*}_{m+1,N}} & {Y^{(m)^*}_{m+1,N}} \\
Y^{(m)}_{m+1,m+1} &X^{(m)}_{m+1,m+1}&Y^{(m)}_{m+1,m+2} & X^{(m)}_{m+1,m+2} &\cdots & Y^{(m)}_{m+1,N} & X^{(m)}_{m+1,N}\\
{X^{(m)^*}_{m+2,m+1}} & {Y^{(m)^*}_{m+2,m+1}} &{X^{(m)^*}_{m+2,m+2}} & {Y^{(m)^*}_{m+2,m+2}} & \cdots  &{X^{(m)^*}_{m+2,N}} & {Y^{(m)^*}_{m+2,N}} \\
Y^{(m)}_{m+2,m+1} &X^{(m)}_{m+2,m+1}&Y^{(m)}_{m+2,m+2} & X^{(m)}_{m+2,m+2} &\cdots & Y^{(m)}_{m+2,N} & X^{(m)}_{m+2,N}\\
\vdots &\vdots &\vdots &\vdots &\ddots &\vdots &\vdots \\
{X^{(m)^*}_{N,m+1}} & {Y^{(m)^*}_{N,m+1}} &{X^{(m)^*}_{N,m+2}} & {Y^{(m)^*}_{N,m+2}} & \cdots  &{X^{(m)^*}_{N,N}} & {Y^{(m)^*}_{N,N}} \\
Y^{(m)}_{N,m+1} &X^{(m)}_{N,m+1}&Y^{(m)}_{N,m+2} & X^{(m)}_{N,m+2} &\cdots & Y^{(m)}_{N,N} & X^{(m)}_{N,N}
\end{pmatrix}.
\ee

After the construction of the reduced density matrix, one can use the standard replica method \cite{Callan:1994py, Holzhey:1994we, Calabrese:2004eu, Calabrese:2009qy} to construct the its $n$-th power in order to work out its trace. This step is basically the same for both of our models which is pictorially explained in Fig.\ref{fig:replica} for $m=1$ and $m=2$ and $N=4$. The replica method here is exactly the same as the well-known procedure for 2d CFTs within the context of spatial entanglement (e.g. see \cite{Calabrese:2009qy}). The only difference is that here we cut along the whole spatial coordinates at $\tau=0$ of those fields which we are not integrating out (see Fig.\ref{fig:replica}).

\begin{figure}
\begin{center}
\begin{tikzpicture}[scale=1.5]
\draw [blue!40!black] (0,0) -- (0,2);
\draw [black] (0.2,0) -- (0.2,2);
\draw [black] (0.4,0) -- (0.4,2);
\draw [black] (0.6,0) -- (0.6,.9);
\draw [black] (0.6,1.1) -- (0.6,2);
\draw [black] (1.5,0) -- (1.5,2);
\draw [black] (1.7,0) -- (1.7,2);
\draw [black] (1.9,0) -- (1.9,2);
\draw [black] (2.1,0) -- (2.1,.9);
\draw [black] (2.1,1.1) -- (2.1,2);
\draw [black] (3,0) -- (3,2);
\draw [black] (3.2,0) -- (3.2,2);
\draw [black] (3.4,0) -- (3.4,2);
\draw [black] (3.6,0) -- (3.6,.9);
\draw [black] (3.6,1.1) -- (3.6,2);
\draw [red] (.6,.9) -- (2.1,1.1);
\draw [red] (2.1,.9) -- (3.6,1.1);
\draw [red,->](3.6,.9) arc (-90:0:.2cm and .2cm);
\draw [red](3.8,1.1) arc (0:90:.2cm and .2cm);
\draw [red, densely dashed,->] (3.6,1.3) -- (2.3,1.3);
\draw [red, densely dashed,->] (2.3,1.3) -- (1,1.3);
\draw [red, densely dashed] (1,1.3) -- (.6,1.1);
\draw [black,->] (-.5,.7)--(-.5,1.3);
\draw [black] (-.65,1) node {$t$};
\draw [black] (0,-.2) node {$4$};
\draw [black] (.2,-.2) node {$3$};
\draw [black] (.4,-.2) node {$2$};
\draw [black] (.6,-.2) node {$1$};
\draw [black] (1.5,-.2) node {$4$};
\draw [black] (1.7,-.2) node {$3$};
\draw [black] (1.9,-.2) node {$2$};
\draw [black] (2.1,-.2) node {$1$};
\draw [black] (3,-.2) node {$4$};
\draw [black] (3.2,-.2) node {$3$};
\draw [black] (3.4,-.2) node {$2$};
\draw [black] (3.6,-.2) node {$1$};
\draw [black] (5.5,0) -- (5.5,2);
\draw [black] (0.2+5.5,0) -- (0.2+5.5,2);
\draw [black] (0.4+5.5,0) -- (0.4+5.5,.9);
\draw [black] (0.4+5.5,1.1) -- (0.4+5.5,2);
\draw [black] (0.6+5.5,0) -- (0.6+5.5,.9);
\draw [black] (0.6+5.5,1.1) -- (0.6+5.5,2);
\draw [black] (1.5+5.5,0) -- (1.5+5.5,2);
\draw [black] (1.7+5.5,0) -- (1.7+5.5,2);
\draw [black] (1.9+5.5,0) -- (1.9+5.5,.9);
\draw [black] (1.9+5.5,1.1) -- (1.9+5.5,2);
\draw [black] (2.1+5.5,0) -- (2.1+5.5,.9);
\draw [black] (2.1+5.5,1.1) -- (2.1+5.5,2);
\draw [black] (3+5.5,0) -- (3+5.5,2);
\draw [black] (3.2+5.5,0) -- (3.2+5.5,2);
\draw [black] (3.4+5.5,0) -- (3.4+5.5,.9);
\draw [black] (3.4+5.5,1.1) -- (3.4+5.5,2);
\draw [black] (3.6+5.5,0) -- (3.6+5.5,.9);
\draw [black] (3.6+5.5,1.1) -- (3.6+5.5,2);
\draw [red] (.6+5.5,.9) -- (2.1+5.5,1.1);
\draw [red] (2.1+5.5,.9) -- (3.6+5.5,1.1);
\draw [red] (.4+5.5,.9) -- (1.9+5.5,1.1);
\draw [red] (1.9+5.5,.9) -- (3.4+5.5,1.1);
\draw [red,->, densely dashed](3.6+5.5,.9) arc (-90:0:.2cm and .2cm);
\draw [red, densely dashed](3.8+5.5,1.1) arc (0:90:.2cm and .2cm);
\draw [red,->, densely dashed](3.4+5.5,.9) arc (-90:0:.25cm and .3cm);
\draw [red, densely dashed](3.65+5.5,1.2) arc (0:90:.2cm and .3cm);
\draw [red, densely dashed,->] (3.6+5.5,1.3) -- (2.3+5.5,1.3);
\draw [red, densely dashed,->] (2.3+5.5,1.3) -- (1+5.5,1.3);
\draw [red, densely dashed,->] (3.4+5.5,1.5) -- (2.3+5.5,1.5);
\draw [red, densely dashed,->] (2.3+5.5,1.5) -- (1.1+5.5,1.5);
\draw [red, densely dashed] (1+5.5,1.3) -- (.6+5.5,1.1);
\draw [red, densely dashed] (1.1+5.5,1.5) -- (.4+5.5,1.1);
\draw [black] (0+5.5,-.2) node {$4$};
\draw [black] (.2+5.5,-.2) node {$3$};
\draw [black] (.4+5.5,-.2) node {$2$};
\draw [black] (.6+5.5,-.2) node {$1$};
\draw [black] (1.5+5.5,-.2) node {$4$};
\draw [black] (1.7+5.5,-.2) node {$3$};
\draw [black] (1.9+5.5,-.2) node {$2$};
\draw [black] (2.1+5.5,-.2) node {$1$};
\draw [black] (3+5.5,-.2) node {$4$};
\draw [black] (3.2+5.5,-.2) node {$3$};
\draw [black] (3.4+5.5,-.2) node {$2$};
\draw [black] (3.6+5.5,-.2) node {$1$};
\end{tikzpicture}
\end{center}
\caption{Replica method for $N=4$ and $n=3$ for $m=1$ (left) and $m=2$ (right). The spatial directions of the field theory are perpendicular to the plane and the vertical lines correspond to the time direction. The numbers under each vertical line corresponds to $i$-th field $\phi_i$.}
\label{fig:replica}
\end{figure}
What remains to do is to start from Eq.\eqref{eq:rhomAmm} and find the trace of the reduced density matrix for general Renyi index $n$ for generic $m$ and $N$.
It is not a hard task, although messy, to see that using replica method one can find
\begin{align}\label{eq:rhoMnN}
\begin{split}
\mathrm{Tr}&\left[\rho_{(N-m)}^n\right]=
\int \mathcal{D}\phi^{(1)}_{m+1}\cdots\mathcal{D}\phi^{(N)}_{m+1}
\mathcal{D}\phi^{(1)}_{m+2}\cdots\mathcal{D}\phi^{(N)}_{m+2}\cdots \cdots \mathcal{D}\phi^{(1)}_{N}\cdots\mathcal{D}\phi^{(N)}_{N}\times\\
&\exp\Bigg\{-\frac{1}{2}\int dx^{d-1}dy^{d-1}
\begin{pmatrix}\phi^{(1)}_{m+1}(x)&\cdots&\phi^{(N)}_{m+1}(x)&\cdots\cdots &\phi^{(1)}_{N}(x)&\cdots&\phi^{(N)}_{N}(x)\end{pmatrix}\cdot \mathcal{M} \cdot
\begin{pmatrix}\phi^{(1)}_{m+1}(y) \\ \vdots \\ \phi^{(N)}_{m+1}(y) \\ \vdots \\ \vdots \\ \phi^{(1)}_{N}(y) \\ \vdots \\ \phi^{(N)}_{N}(y)\end{pmatrix} \Bigg\},
\end{split}
\end{align}
where the matrix $\mathcal{M}$ is a $n(N-m)\times n(N-m)$ square matrix and is defined in terms of $\mathcal{M}_{m,m'}$ blocks as
\be
\mathcal{M}=
\begin{pmatrix}
\mathcal{M}_{m+1,m+1} & \mathcal{M}_{m+1,m+2} & \cdots  &\mathcal{M}_{m+1,N} \\
\mathcal{M}_{m+2,m+1} & \mathcal{M}_{m+2,m+2} & \cdots  &\mathcal{M}_{m+2,N} \\
\vdots &\vdots &\ddots &\vdots\\
\mathcal{M}_{N,m+1} & \mathcal{M}_{N,m+2} & \cdots  &\mathcal{M}_{N,N}
\end{pmatrix},
\ee
and the blocks $\mathcal{M}_{m,m'}$ are $n\times n$ square matrices given by
\be
\mathcal{M}_{m,m'}=
\begin{pmatrix}
\widetilde{X}_{m,m'} & Y_{m,m'} & 0 & \cdots  &Y_{m',m}\\
Y_{m',m} & \widetilde{X}_{m,m'} & Y_{m,m'}  & \cdots & 0 \\
0 & Y_{m',m} & \widetilde{X}_{m,m'}  & \cdots & 0 \\
\vdots &\vdots &\vdots  & \ddots & \vdots \\
 0 & 0  &0 & \widetilde{X}_{m,m'} & Y_{m,m'}\\
Y_{m,m'} & 0 & 0 & Y_{m',m} & \widetilde{X}_{m,m'}
\end{pmatrix}.
\ee


If we calculate the determinate of $\mathcal{M}$ we are done. This would be a much simpler task if we consider the explicit values of $G_{ij}$'s for the infinite-range model. To do so the key point is the existence of an orthogonal transformation which results in a diagonal model (free scalar fields) as was explained in section \ref{sec:model1} and specifically in Eq.\eqref{eq:Sdiag}. In the diagonal basis the ground state wave functional up to a normalization constant becomes
\be
\Psi[\phi'_1,\cdots,\phi'_N]=\exp\left\{-\frac{1}{2}\int dx^{d-1}dy^{d-1}\,W(x,y)\left[\sum_{i=1}^N A_i\phi'_i(x)\phi'_i(y)\right]\right\},
\ee
where $A_i$'s are given in Eq.\eqref{eq:A1} and $W(x,y)$ is given by
\be
W(x,y)=\frac{1}{V}\sum_k|k|e^{ik(x-y)},
\ee
where $V$ is the $(d-1)$-dimensional volume which the field theory is defined on. Since we have applied an orthogonal transformation between $\left\{\phi_1,\cdots,\phi_N\right\}$ and $\left\{\phi'_1,\cdots,\phi'_N\right\}$ basis, the physical state is unaffected, i.e. 
$$\Psi[\phi'_1,\cdots,\phi'_N]=\Psi[\phi_1,\cdots,\phi_N],$$
and we can rewrite the ground state in terms of $\left\{\phi_1,\cdots,\phi_N\right\}$ basis as
\be
\Psi[\phi_1,\cdots,\phi_N]=\exp\left\{-\frac{1}{2}\int dx^{d-1}dy^{d-1}\,W(x,y)\left[\sum_{i,j=1}^NG_{ij}\phi_i(x)\phi_j(y)\right]\right\},
\ee
where $G_{ij}$'s for this model are given by
\be\label{eq:Gij1A}
G=\frac{1}{2}
\begin{pmatrix}
            2 & \lambda & \lambda & \cdots &\lambda   \\
            \lambda &2&\lambda&\cdots &\lambda \\
             \lambda&\lambda&2&\cdots &\lambda \\
            \vdots &\vdots &\vdots &\ddots &\vdots \\
             \lambda&\lambda&\lambda&\cdots &2 \\
             \end{pmatrix}.
\ee
Using these explicit expressions and working out the trace of the reduced density matrix first for $m=1$ and generic $N$, by induction one can easily find that 
\be\label{eq:TrrhonIA2}
\mathrm{Tr}\left[\rho_{(1)}^n\right]=\mathcal{N}\prod_{i}\prod_{r=1}^n\left[1+\frac{(N-1)\lambda^2}{(N-1)\lambda^2-4\lambda(N-1)+4(\lambda-2)}\cos\left(\frac{2\pi r}{n}\right)\right].
\ee
Now we are done with the $m=1$ case. Generalizing to $m>1$ is not a hard task because of a simple structure in the reduced density matrix. Since the structure of the reduced density matrix only depends on $(N-m)$ rather than $m$ and $N$ itself, we are almost done since we already have calculated $m=1$ for generic $N$. Again by induction one can generalize the above result for general $m$ which is
\be\label{eq:TrrhonIAA}
\mathrm{Tr}\left[\rho_{(m)}^n\right]=\mathcal{N}\prod_{i}\prod_{r=1}^n\left[1+\frac{4(N-m)Y(m)}{4(N-m)Y(m)+(N-m)\lambda+2-\lambda}\cos\left(\frac{2\pi r}{n}\right)\right],
\ee
where $Y(m)$ (not to be confused with the $Y$ elements of matrix $\mathcal{M}$) is defined as
\be
Y(m)=-\frac{1}{4}\left(\frac{\lambda}{2}\right)^2\cdot\frac{2m}{2+(m-1)\lambda}.
\ee
\vspace{3mm}

\subsection{Nearest-Neighbour Model}\label{sec:ANN}
The logical steps for this model is the same as that we have discussed in the previous subsection. We may start from Eq.\eqref{eq:rhomAm} for this model. In comparison with the infinite-range model, this model has much fewer symmetries which makes it harder to push this calculation as general as we did for the infinite-range model. Since we are interested in the case where the strength of interactions between interacting fields is equal, we will restrict our analysis for equal off diagonal values of $G_{ij}$ which we denote by $G_{ij}\equiv G$ for $i\neq j$, and also $G_{ii}\equiv G_d$. For such a case one can perform $m$ number of Gaussian integrals to arrive at the general reduced density matrix in the form of Eq.\eqref{eq:rhomAmm} with
\be
M(m,N)=\begin{pmatrix}
G_d^*-{X_m^*}^2 & |X_m|^2 & G^* & 0 & \cdots & {Y_m^*}^2 & |Y_m|^2 \\
|X_m|^2 & G_d-X_m^2 & 0 & G & \cdots &|Y_m|^2 & Y_m^2 \\
G^* & 0 &G_d^* & 0 &\cdots & 0 & 0\\
0 & 0 & 0 & G_d & \cdots  &0 & 0 \\
\vdots &\vdots &\vdots &\vdots &\ddots &\vdots &\vdots \\
{Y_m^*}^2 & |Y_m|^2 &0 &0 & \cdots  &G_d^*-{X_m^*}^2 & |X_m|^2 \\
|Y_m|^2 & Y_m^2  & 0 &0 &\cdots & |X_m|^2 & G_d-X_m^2  
\end{pmatrix}.
\ee
where
\be
X_m=G\left[\frac{1}{4Z_m}\right]^{\frac{1}{2}}\;\;\;\;,\;\;\;\;
Y_m=G\left[\frac{{\widetilde{G}}^{m-1}}{(-2)^m\prod_{i=1}^mZ_i}\right]^{\frac{1}{2}}\;\;\;\;,\;\;\;\;
Z_m=Z_1-\frac{\widetilde{G}^2}{4Z_{m-1}},
\ee
and  $Z_1=\widetilde{G_d}$. Note that the above general form is correct for $m>1$, for the case of $m=1$ there is an extra factor of 2 in the denominator of all components represented in terms of $Y_m$. The reader should note that these $Y_m$ and $Z_m$ functions are not to be confused with the functions with $Y(m)$, $Y_d(m)$ and $Z(m)$ which appear in the final result as functions of the coupling which is given in Eq.\eqref{eq:TrrhonII} and Eq.\eqref{eq:TrrhonII2}.

Now we can work out the counterpart of Eq.\eqref{eq:rhoMnN} in this model. Here the form of $\mathcal{M}$ is more complicated and is given as follows
\be\label{eq:CM2}
\mathcal{M}=
\begin{pmatrix}
\mathcal{M}^X & \mathcal{M}^G & 0 & 0 & \cdots  &\mathcal{M}^Y \\
\mathcal{M}^G & \mathcal{M}^{G_d} & \mathcal{M}^G & 0& \cdots  &0 \\
0 & \mathcal{M}^{G} & \mathcal{M}^{G_d} & \mathcal{M}^G & \cdots  & 0 \\
\vdots &\vdots &\vdots &\vdots &\ddots &\vdots\\
0 & 0 & \cdots & \mathcal{M}^{G} & \mathcal{M}^{G_d}& \mathcal{M}^{G}\\
\mathcal{M}^Y & 0 & \cdots & 0 & \mathcal{M}^{G} &\mathcal{M}^X
\end{pmatrix},
\ee
where again the blocks $\mathcal{M}^i$ are $n\times n$ square matrices given by
\begin{align}
\begin{split}
\mathcal{M}^X&=
\begin{pmatrix}
\widetilde{G_d}-\widetilde{X_m^2} & |X_m|^2 & 0 & \cdots  &|X_m|^2\\
|X_m|^2 & \widetilde{G_d}-\widetilde{X_m^2} & |X_m|^2  & \cdots & 0 \\
0 & |X_m|^2 & \widetilde{G_d}-\widetilde{X_m^2}  & \cdots & 0 \\
\vdots &\vdots &\vdots  & \ddots & \vdots \\
 0 & 0  &|X_m|^2 & \widetilde{G_d}-\widetilde{X_m^2} & |X_m|^2\\
|X_m|^2 & 0 & 0 & |X_m|^2 & \widetilde{G_d}-\widetilde{X_m^2}
\end{pmatrix}\\
\mathcal{M}^Y&=
\begin{pmatrix}
\widetilde{Y_m^2} & |Y_m|^2 & 0 & \cdots  &|Y_m|^2\\
|Y_m|^2 & \widetilde{Y_m^2} & |Y_m|^2  & \cdots & 0 \\
0 & |Y_m|^2 & \widetilde{Y_m^2}  & \cdots & 0 \\
\vdots &\vdots &\vdots  & \ddots & \vdots \\
 0 & 0  &|Y_m|^2 & \widetilde{Y_m^2} & |Y_m|^2\\
|Y_m|^2 & 0 & 0 & |Y_m|^2 & \widetilde{Y_m^2}
\end{pmatrix}
\end{split}
\end{align}
and $M^G=\mathrm{diag}\{G,\cdots,G\}$ and $M^{G_d}=\mathrm{diag}\{G_d,\cdots,G_d\}$.

Now we are equipped with $\mathrm{Tr}\left[\rho_{(N-m)}^n\right]$ for the nearest-neighbour model and what remains is to plug in the corresponding $G_{ij}$ which was given in Eq.\eqref{eq:Gij2} and work out the determinant of $\mathcal{M}$ given in Eq.\eqref{eq:CM2}. This step is of course more messy than the case of infinite-range model because of a technical subtlety. Here in contrast with the infinite-range model, when we increase $m$ and $N$, the degree of the polynomials appearing in the expression of $\mathrm{det}[\mathcal{M}]$ also increases. The key point to bring these expressions back into control is to factor them in terms of their roots, which generally take the form of $\lambda^{-1}=\cos\left[w(m,N)\pi\right]$ with different $w(m,N)$ functions.
Following such a process will lead to Eq.\eqref{eq:TrrhonII}. Note that the functions $X$ and $Y$ used here has nothing to do with the functions given in the final result Eq.\eqref{eq:TrrhonII}.

\end{document}